\documentclass[10pt,conference]{IEEEtran}
\usepackage{amsmath,amssymb,amsfonts}
\usepackage{algorithmic}
\usepackage{textcomp}

\usepackage{xspace}

\usepackage{tikz}
\usepackage{xcolor}
\newcommand*\circled[1]{\tikz[baseline=(char.base)]{
            \node[shape=circle,fill,inner sep=1.5pt] (char) {\textcolor{white}{#1}};}}

\usepackage{mathptmx} 

\usepackage{fancyhdr}
\usepackage[normalem]{ulem}
\usepackage[sort,nocompress]{cite}

\usepackage{multirow}
\usepackage{makecell}
\usepackage{array}
\usepackage{adjustbox}
\usepackage{xcolor}
\usepackage{tikz}
\usepackage{svg}
\usepackage{comment}
\svgsetup{inkscapelatex=false}

\usepackage{graphicx}
\usepackage{caption}
\usepackage{booktabs}
\usepackage{varwidth}
\newsavebox\tmpbox

\usepackage[bookmarks=true,breaklinks=true,colorlinks,citecolor=blue,linkcolor=blue,urlcolor=blue]{hyperref}

\newcommand{\hpcayear}{2024}

\newcommand{\ie}{\textit{i.e.}}
\newcommand{\eg}{\textit{e.g.}}
\newcommand{\cf}{\textit{cf.}\xspace}
\def\arch{SPADE\xspace}

\newcommand{\hpcasubmissionnumber}{NaN}
\title{SPADE: Sparse Pillar-based 3D Object Detection Accelerator for Autonomous Driving}

\def\hpcacameraready{} 

\newcommand\hpcaauthors{Minjae Lee\textsuperscript{$\dagger$}, Seongmin Park\textsuperscript{$\dagger$}, Hyungmin Kim\textsuperscript{$\dagger$}, Minyong Yoon\textsuperscript{$\dagger$}, Janghwan Lee\textsuperscript{$\dagger$},\\ Jun Won Choi\textsuperscript{$\dagger$}, Nam Sung Kim\textsuperscript{$\ddagger$}, Mingu Kang\textsuperscript{$\S$}, Jungwook Choi\textsuperscript{$\dagger$*}}
\newcommand\hpcaaffiliation{\textsuperscript{$\dagger$}Hanyang University, \textsuperscript{$\ddagger$}University of Illinois at Urbana-Champaign, \textsuperscript{$\S$}University of California, San Diego}
\newcommand\hpcaemail{\{lmj4666, skstjdals, kong4274, ycivil93, hwanii0288, junwchoi, choij\}@hanyang.ac.kr,\\ nskim@illinois.edu, mingu@ucsd.edu} 

\author{
  \ifdefined\hpcacameraready
    \IEEEauthorblockN{\hpcaauthors{}}
      \IEEEauthorblockA{
        \hpcaaffiliation{} \\
        \hpcaemail{}
      }
  \else
    \IEEEauthorblockN{\normalsize{HPCA \hpcayear{} Submission
      \textbf{\#\hpcasubmissionnumber{}}} \\
      \IEEEauthorblockA{
        Confidential Draft \\
        Do NOT Distribute!!
      }
    }
  \fi 
}

\fancypagestyle{camerareadyfirstpage}{%
  \fancyhead{}
  
  \fancyhead[C]{
    \ifdefined\aeopen
    \parbox[][12mm][t]{13.5cm}{\hpcayear{} IEEE International Symposium on High-Performance Computer Architecture (HPCA)}    
    \else
      \ifdefined\aereviewed
      \parbox[][12mm][t]{13.5cm}{\hpcayear{} IEEE International Symposium on High-Performance Computer Architecture (HPCA)}
      \else
      \ifdefined\aereproduced
      \parbox[][12mm][t]{13.5cm}{\hpcayear{} IEEE International Symposium on High-Performance Computer Architecture (HPCA)}
      \else
      \parbox[][0mm][t]{13.5cm}{\hpcayear{} IEEE International Symposium on High-Performance Computer Architecture (HPCA)}
    \fi 
    \fi 
    \fi 
    \ifdefined\aeopen 
      \includegraphics[width=12mm,height=12mm]{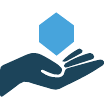}
    \fi 
    \ifdefined\aereviewed
      \includegraphics[width=12mm,height=12mm]{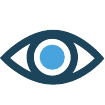}
    \fi 
    \ifdefined\aereproduced
      \includegraphics[width=12mm,height=12mm]{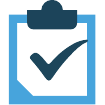}
    \fi
  }
  \fancyfoot[C]{}
}
\begin{document}
\maketitle

\ifdefined\hpcacameraready 
  \thispagestyle{camerareadyfirstpage}
  \pagestyle{empty}
\else
  \thispagestyle{plain}
  \pagestyle{plain}
\fi

\newcommand{\hpcaheight}{0mm}
\ifdefined\eaopen
\renewcommand{\hpcaheight}{12mm}
\fi
  

\begin{abstract}

3D object detection using point cloud (PC) data is essential for perception pipelines of autonomous driving, where efficient encoding is key to meeting stringent resource and latency requirements. PointPillars, a widely adopted bird's-eye view (BEV) encoding, aggregates 3D point cloud data into 2D \textit{pillars} for fast and accurate 3D object detection. However, the state-of-the-art methods employing PointPillars overlook the inherent sparsity of pillar encoding where only a valid pillar is encoded with a \textit{vector} of channel elements, missing opportunities for significant computational reduction. Meanwhile, current sparse convolution accelerators are designed to handle only element-wise activation sparsity and do not effectively address the \textit{vector sparsity} imposed by pillar encoding.

In this paper, we propose \arch, an algorithm-hardware co-design strategy to maximize \textit{vector sparsity} in pillar-based 3D object detection and accelerate \textit{vector-sparse convolution} commensurate with the improved sparsity. \arch consists of three components: (1) a \textit{dynamic vector pruning algorithm} balancing accuracy and computation savings from vector sparsity, (2) a \textit{sparse coordinate management hardware} transforming 2D systolic array into a vector-sparse convolution accelerator, and (3) \textit{sparsity-aware dataflow optimization} tailoring sparse convolution schedules for hardware efficiency. Taped-out with a commercial technology, \arch saves the amount of computation by 36.3--89.2\%  for representative 3D object detection networks and benchmarks, leading to 
1.3--10.9$\times$ speedup and 1.5--12.6$\times$ energy savings compared to the ideal dense accelerator design. These sparsity-proportional performance gains equate to 4.1--28.8$\times$ speedup and 90.2--372.3$\times$ energy savings compared to the counterpart server and edge platforms. 

\end{abstract}
\footnotetext[1]{Corresponding author.}

\section{Introduction}
\label{sec:intro}

The importance of autonomous driving has shifted from driving convenience to enhancing safety and traffic flow~\cite{arnold2019survey}. Advanced autonomous driving technologies depend on robust perception systems to transform sensory data into semantic information, such as identifying and locating road agents (\eg, vehicles and pedestrians). 3D object detection, a vital component of automotive perception systems, overcomes limitations of 2D object detection by providing depth information in world coordinates~\cite{arnold2019survey, mao20223d}. Point cloud (PC) 3D object detection, in particular, captures obstacle distances using LiDAR sensors, offering essential cues for 3D-scene perception while demanding intricate handling of unstructured and sparse PC data. 

Fast PC-based 3D object detection is crucial for the safe and efficient operation of autonomous driving systems~\cite{mao20223d}. Typically, \textit{real-time} perception involves processing sensory data at video rates (\ie, 10-100~FPS). However, since rapid responses to potential hazards or collisions depend on perception information, processing speeds \textit{above 100~FPS} are desirable for allowing vehicles to make swift, informed decisions in response to changing road conditions and obstacles. Moreover, pushing the boundaries of real-time processing would contribute to a smooth and responsive driving experience, which is critical for the public's acceptance of autonomous vehicles and helps manufacturers and technology providers distinguish themselves in a competitive market.

In response to the growing demand for real-time PC-based 3D object detection in autonomous driving systems, researchers have concentrated on developing faster and more accurate techniques, transitioning from point-based methods \cite{qi2018frustum, shi2019pointrcnn, yang2019std} to grid-based approaches \cite{zhou2018voxelnet,yan2018second,lang2019pointpillars} to enhance performance and efficiency for real-time applications. The grid-based 3D object detection quantizes point clouds into 3D grid voxels to derive features through convolution operations. Specifically, PointPillars~\cite{lang2019pointpillars}, the foundational grid-based approach simplifies neighborhood search with two techniques: (1) a bird's-eye-view (BEV) encoding that aggregates voxels vertically into \textit{pillars}, or sparse vectors of channel elements, and (2) the densification of sparse pillars into a pseudo-image for GPU-friendly feature extraction using 2D convolution (Conv2D). As a result, PointPillars has become a prominent backbone for fast 3D object detection~\cite{wang2020pillar, zhou2020end, sun2020scalability, li2021lidar,yin2021center, shi2022pillarnet}. However, its densification approach overlooks the inherent sparsity of pillar encoding, which usually represents only $5\%$ of the dense pseudo-image, resulting in unnecessary computation. 

\begin{figure*}[htbp]
\begin{center}
\centerline{
\includegraphics[width=0.75\linewidth]{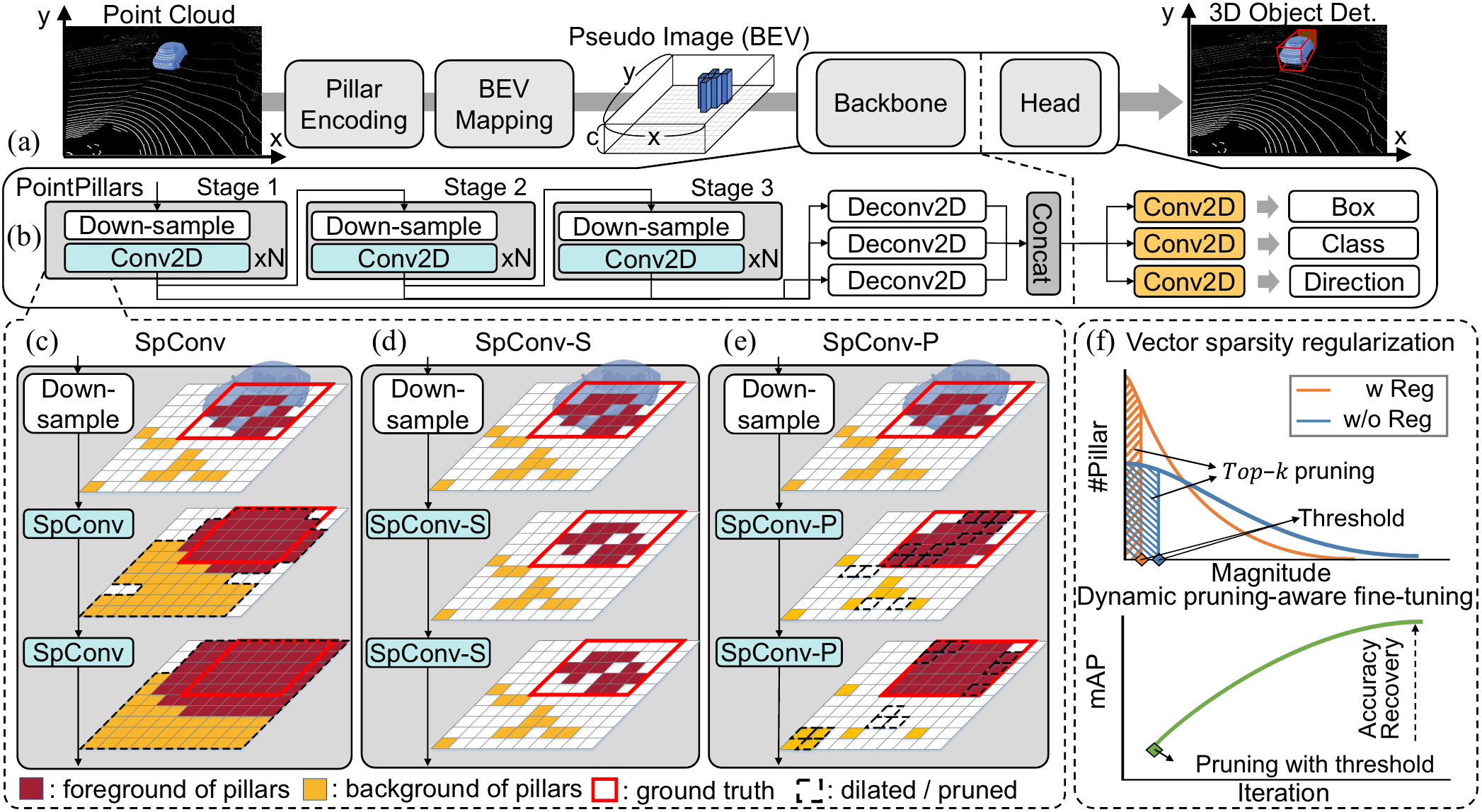}
}
\caption{Overview of pillar-based 3D object detection: (a) model structure, (b) feature extraction steps (backbone and head). Comparison of receptive fields (Stage 1) for various sparse convolution
operations: (c) Spconv, (d) Spconv-S, and (e) SpConv-P. (f) Dynamic vector pruning for SpConv-P: vector sparsity regularization and dynamic pruning-aware fine-tuning.}
\label{fig:fig2}
\end{center}
\vspace{-0.7cm}
\end{figure*}

Despite the potential computational benefits, processing sparse pillars using existing accelerators presents challenges due to their \textit{vector sparsity}, characterized by consecutive zeros across all channels for each inactive pillar. This pattern is distinct from that handled by traditional sparse Conv2D accelerators~\cite{parashar2017scnn, wang2021dstc, yang2021s2, shabani2023hirac}, which are geared towards managing element-wise activation sparsity where zeros occur randomly due to the ReLU operation. These accelerators use an outer-product methodology for sparse convolution, which accumulates partial-sums in an output-stationary manner to map the irregular locations of non-zero input activations to the output coordinates. However, this irregular input-output mapping can incur significant overhead for sparse pillars because of workload imbalances and bank conflicts during partial-sum aggregation.

An alternative approach to handle vector sparsity is to adopt voxel-sparse convolution~\cite{zhou2018voxelnet,yan2018second}, which explicitly computes an input-output mapping for active pillars (or non-zero vectors). However, this method also results in a significant overhead because it necessitates hashing or sorting to identify the coordinates of related input and output in irregular locations. For instance, \cite{vedder2021sparse} employed submanifold sparse convolution~\cite{graham2017submanifold} for sparse processing of pillar-based feature extraction but only achieved marginal speedup due to the overhead associated with coordinate searching. Past work has suggested approximate search methods to lower mapping costs \cite{xu2019tigris,feng2020mesorasi,feng2022crescent} or proposed custom sorting and cache units for accelerated search \cite{lin2021pointacc}. Nevertheless, these acceleration techniques lead to substantial overhead in input-output mapping and cache misses for vector sparsity, a topic we will discuss further in Sec.~\ref{sec:arch}.

In this work, we propose \arch (\textbf{S}parse \textbf{P}illar-b\textbf{a}sed 3D object \textbf{De}tection accelerator) with an algorithm-hardware co-design to maximize \textit{vector sparsity} in pillar-based 3D object detection and accelerates \textit{vector-sparse convolution} to attain speedup granted by the sparsified computation. \arch consists of three components. First, we develop a new \textit{dynamic vector pruning algorithm} to explore the pareto-optimal trade-off between accuracy and sparsity in 3D object detection models (\S\ref{subsec:motivation}). Second, we design a \textit{sparse coordinate management hardware}, transforming a conventional 2D systolic array into a vector-sparse convolution accelerator featured with linear complexity input-output mapping (\S\ref{subsec:arch_rule-gen}) and conflict-free scatter-gather (\S\ref{subsec:arch_g-s}). Third, we devise \textit{sparsity-aware dataflow optimization} techniques that dynamically adjust sparse convolution schedules to optimize hardware utilization (\S\ref{subsec:arch_g-s-arch}). The resulting \arch accelerator offers sparsity-proportional speedups using minimal additional hardware resources compared to dense convolution counterparts.

We tape out a \arch chip with a commercial technology and evaluate \arch's performance on several popular 3D object detection networks such as PointPillar~\cite{lang2019pointpillars}, CenterPoint~\cite{yin2021center}, and PillarNet~\cite{shi2022pillarnet}, as well as widely-used benchmarks like KITTI~\cite{geiger2012kitti} and nuScenes~\cite{caesar2020nuscenes}. \arch's dynamic vector-pruning algorithm saves the amount of computation by 36.3--89.2\%, leading to 
1.3--10.9$\times$ speedup and 1.5--12.6$\times$ energy savings compared to the ideal dense accelerator design, thanks to high sustained hardware utilization by dataflow optimization.
These performance enhancements, directly proportional to sparsity, equate to 4.1--28.8$\times$ speedup and 90.2--372.3$\times$ energy savings compared to counterpart server-grade GPU (NVIDIA RTX 2080Ti) and advanced edge computing device (NVIDIA Jetson NX). It reached a record-breaking speed of 500~FPS with minimal accuracy loss (mAP -0.4\%), demonstrating its effectiveness and potential for real-time 3D object detection on edge devices, ideal for latency-sensitive applications like autonomous driving.


\section{Pillar-Based 3D Object Detection: \\ Algorithm, Optimization, and Challenge}
\label{sec:background}

\subsection{Pillar-Based 3D Object Detection}
\label{subsec:3d-obj-det}

A point cloud (PC) is a collection of points in 3D space, represented as $P = \{p_k\} = \{(c_k, r_k)\}$, where $c_k = (x_k, y_k, z_k)$ denotes the 3D coordinate of the $k$'th point, and $r_k$ is its feature vector. With the widespread availability of LiDAR devices for point cloud acquisition and the development of deep learning algorithms that can effectively extract semantic information, point clouds have become critical components of intelligent systems for autonomous vehicles. Specifically, PC-based 3D object detection has become a standard pipeline, providing important cues for 3D scene perception by capturing the distance of obstacles~\cite{arnold2019survey, mao20223d}.

As a promising real-time PC-based 3D object detection, the pillar-based approaches involve pillar encoding projecting PC data into 2D space for feature extraction~\cite{wang2020pillar, sun2020scalability, shi2022pillarnet}. As shown in Fig.~\ref{fig:fig2}(a), the sparse pillars represent bird's-eye view (BEV) features derived from points within discretized region $X\times Y$ binning along $z$ dimension. During pillar encoding, $C$-element channels of feature vectors (\ie, pillars) are computed using a multi-layer perceptron (PointNet~\cite{qi2017pointnet}). PointPillars~\cite{lang2019pointpillars} \textit{densified} these pillars into a dense pseudo-image of size $C\times X\times Y$ for GPU-friendly processing. However, due to point cloud sparsity, the corresponding active pillars are also inherently sparse in a 2D grid; \eg, 97\% of densified pillars are zero-vectors reported in \cite{lang2019pointpillars}.

Fig.~\ref{fig:fig2}(b) illustrates the details of PointPillars' feature extraction consisting of a backbone and head. The backbone consists of multiple stages of convolutions led by downsampling layers (\ie~stride=2) for the increased receptive fields. The outcome of all these stages is deconvoluted and then concatenated for box, class, and direction prediction in the head. Recent variations also exist; CenterPoint~\cite{yin2021center} incorporates center-based prediction heads while PillarNet~\cite{shi2022pillarnet} strengthens pillar encoding with additional sparse convolution layers in front. We target these recent pillar-based 3D object detection models for performance improvement.

\subsection{Sparse PointPillars}
\label{subsec:motivation}

\noindent \textbf{Sparse Convolution for Accelerating PointPillars. }
While pillar-based 3D object detection benefits from dense pillar encoding on GPUs, the inherent sparsity of the application can lead to substantial computational redundancy. Sparse convolution has been proposed to reduce feature map size and enhance computational efficiency in object detection, but a comprehensive understanding of various sparse convolution operations and their \textit{sparsity-accuracy trade-off} is crucial for optimizing sparsity. Fig.~\ref{fig:fig2}(c-e) illustrates candidate sparse convolution operations. Standard sparse convolution (SpConv), which applies convolution only to active pillars, can increase active pillar count due to dilation (Fig.~\ref{fig:fig2}(c)). In contrast, submanifold convolution (SpConv-S) restricts dilation to preserve post-convolution sparsity (Fig.~\ref{fig:fig2}(d)). Controlling dilation is essential for balancing accuracy and sparsity. While indiscriminate dilation can lead to unnecessary computation, dilation of important pillars expands their receptive fields, enhancing object detection. For example, \cite{vedder2021sparse} utilized only SpConv-S to avoid dilation but suffered significant accuracy degradation. Therefore, a more sophisticated dilation control is needed to retain sparsity while conserving accuracy.

\noindent \textbf{Dynamic Vector Pruning. }
In addressing the accuracy-sparsity trade-off, we propose a novel sparse convolution based on \textit{dynamic vector pruning} (SpConv-P). As shown in Fig.~\ref{fig:fig2}(e), SpConv-P permits dilation across layers while dynamically pruning nonessential pillars, thereby maintaining sparsity. Its central innovation lies in the dynamic identification and pruning of unimportant background pillars. This is achieved by adopting \textit{vector sparsity regularization} and \textit{dynamic pruning-aware fine-tuning}, techniques typically used in weight pruning but newly applied to dynamic activation pruning (Fig.~\ref{fig:fig2}(f)). During training, we incorporate loss terms that regulate pillar magnitude across channels, motivated by Group Lasso~\cite{wen2016learning} but distinguished itself by dynamically driving the magnitude of unimportant pillars in varying locations towards zero. Furthermore, to avoid the difficulty of on-the-fly adjustment of the pruning thresholds, we fine-tune the model with Top-K pruning per layer to robustify it against the user-specified activation sparsity. After fine-tuning, representative pruning thresholds for the target activation sparsity can be retrieved for inference. Our experiments demonstrate that the proposed dynamic vector pruning consistently delivers robust accuracy while satisfying the targeted sparsity range.

\begin{table}[]
\caption{Sparse pillar-based 3D object detection.}
{\scriptsize
\begin{center}
\begin{tabular}{@{\hspace{0.1cm}}c@{\hspace{0.10cm}}|@{\hspace{0.10cm}}c@{\hspace{0.10cm}}|@{\hspace{0.10cm}}c@{\hspace{0.10cm}}|@{\hspace{0.10cm}}c@{\hspace{0.10cm}}|@{\hspace{0.10cm}}c@{\hspace{0.10cm}}|@{\hspace{0.10cm}}c@{\hspace{0.10cm}}|@{\hspace{0.10cm}}c@{\hspace{0.10cm}}|@{\hspace{0.10cm}}c@{\hspace{0.10cm}}}
\hline
Dataset & Model & \multicolumn{2}{c|}{Backbone} & Head & \begin{tabular}[c]{@{}c@{}}Avg.GOPs \\ 
 $\left[\begin{array}{@{}c@{}}\text{Sparsity}\end{array}\right]$ \end{tabular} & \begin{tabular}[c]{@{}c@{}}mAP\\ (BEV)\end{tabular} & \begin{tabular}[c]{@{}c@{}}mAP\\ (3D)\end{tabular} \\ \hline
 & PP~\cite{lang2019pointpillars} & \multicolumn{2}{c|}{Conv2D} & Conv2D & \begin{tabular}[c]{@{}c@{}}46.43 \\ $\left[\begin{array}{@{}c@{}}\text{0.0\%}\end{array}\right]$\end{tabular} & 87.42 & 77.31 \\
\multirow{2}{*}{KITTI} & SPP1 & \multicolumn{2}{c@{\hspace{0.15cm}}|}{SpConv} & Conv2D & \begin{tabular}[c]{@{}c@{}}20.33 \\ $\left[\begin{array}{@{}c@{}}\text{56.2\%}\end{array}\right]$\end{tabular} & 87.34 & 77.16 \\
 & SPP2 & \multicolumn{2}{c|}{SpConv-P} & Conv2D & \begin{tabular}[c]{@{}c@{}}12.30\\ $\left[\begin{array}{@{}c@{}}\text{73.5\%}\end{array}\right]$\end{tabular} & 86.99 & 76.94 \\
 & SPP3 & \multicolumn{2}{c|}{SpConv-S} & Conv2D & \begin{tabular}[c]{@{}c@{}}5.01\\ $\left[\begin{array}{@{}c@{}}\text{89.2\%}\end{array}\right]$\end{tabular} & 83.11 & 74.91 \\ \hline
Dataset & Model & \multicolumn{2}{c|}{Backbone} & Head & Avg.GOPs & mAP & NDS \\ \hline
\multirow{12}{*}{nuScenes} & CP~\cite{yin2021center} & \multicolumn{2}{c|}{Conv2D} & Conv2D & \begin{tabular}[c]{@{}c@{}}63.99\\ $\left[\begin{array}{@{}c@{}}\text{0.0\%}\end{array}\right]$\end{tabular} & 50.79 & 60.55 \\
 & SCP1 & \multicolumn{2}{c|}{SpConv} & Conv2D & \begin{tabular}[c]{@{}c@{}}40.76\\ $\left[\begin{array}{@{}c@{}}\text{36.3\%}\end{array}\right]$\end{tabular} & 50.54 & 60.57 \\
 & SCP2 & \multicolumn{2}{c|}{SpConv-P} & SpConv-P & \begin{tabular}[c]{@{}c@{}}24.77\\ $\left[\begin{array}{@{}c@{}}\text{61.3\%}\end{array}\right]$\end{tabular} & 50.12 & 60.42 \\
 & SCP3 & \multicolumn{2}{c|}{SpConv-S} & SpConv-P & \begin{tabular}[c]{@{}c@{}}13.60\\ $\left[\begin{array}{@{}c@{}}\text{78.8\%}\end{array}\right]$\end{tabular} & 47.78 & 58.97 \\ \cline{2-8}
 & {Model} & {Encoder} & {Backbone} & {Head} & {Avg.GOPs} & {mAP} & {NDS} \\ \cline{2-8} 
 & {PN (Dense)} & {Conv2D} & {Conv2D} & {Conv2D} & \begin{tabular}[c]{@{}c@{}}{596.51} \\ $\left[\begin{array}{@{}c@{}}{\text{0.0\%}}\end{array}\right]$\end{tabular} & {59.58} & {66.95} \\
 & {PN~\cite{shi2022pillarnet}} & {SpConv-S} & {Conv2D} & {Conv2D} & \begin{tabular}[c]{@{}c@{}}{284.09} \\ $\left[\begin{array}{@{}c@{}}{\text{52.4\%}}\end{array}\right]$\end{tabular} & {59.58} & {66.95} \\
 & {SPN} & {SpConv-S} & {SpConv-S} & {Conv2D} & \begin{tabular}[c]{@{}c@{}}{160.27}\\ $\left[\begin{array}{@{}c@{}}{\text{73.1\%}}\end{array}\right]$\end{tabular} & {57.92} & {66.33} \\ \hline
\end{tabular}
\vspace{-0.7cm}

\end{center}
}
\label{tab:cp_pn_perf}
\end{table}

\begin{figure*}[]
\begin{center}
\centerline{
\includegraphics[width=0.97\linewidth]{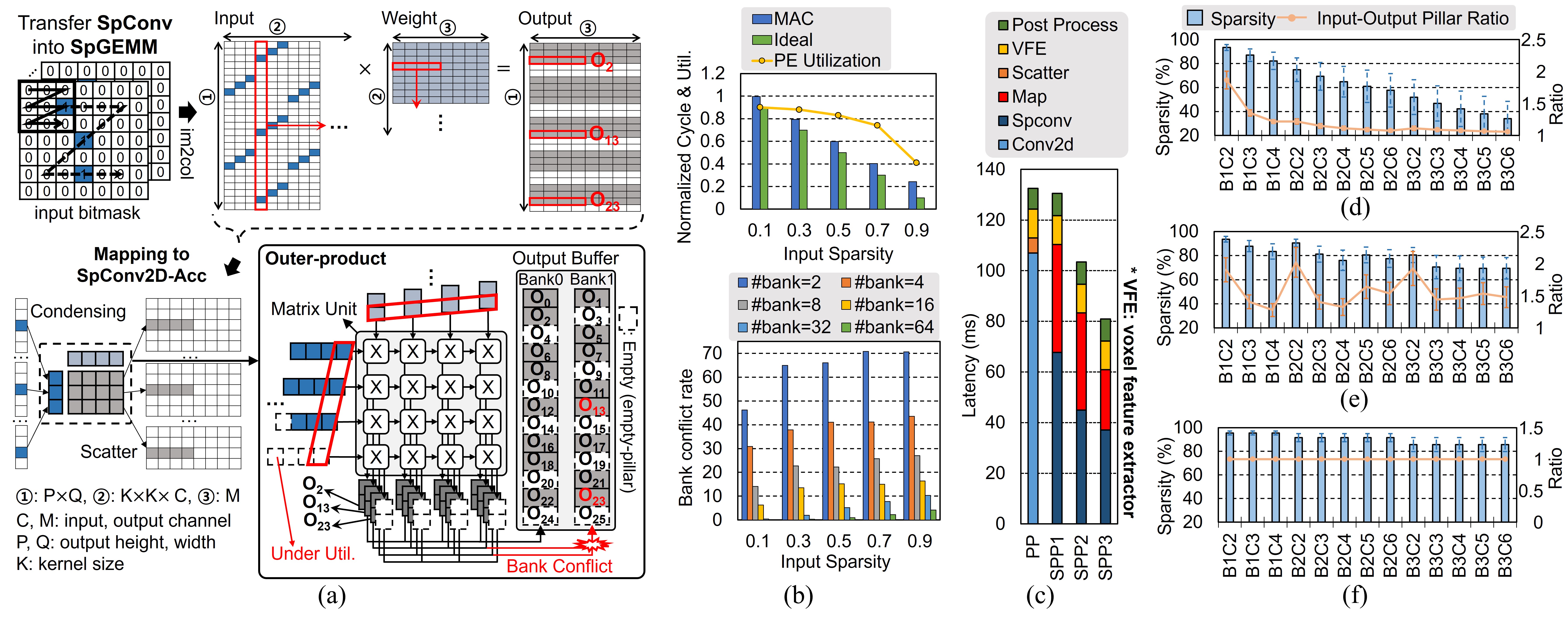}
}
\caption{Conventional sparse Conv2D accelerator for sparse pillars: (a) dataflow, (b) inefficiency (utilization, bank conflict rate). (c) Latency breakdown of PointPillars. Sparsity characteristics of sparse convolution variants: (d) SPP1, (e) SPP2, and (f) SPP3.}
\label{fig:latency-breakdown}
\end{center}
\vspace{-0.6cm}
\end{figure*}

\noindent \textbf{Sparsity Exploration. } 
Our innovative method, SpConv-P, bridges the gap between SpConv and SpConv-S, enabling the exploration of accuracy-sparsity trade-off. By applying different sparse convolution configurations, we built seven sparse pillar-based models from the prominent pillar-based 3D object detection models: three sparse models (SPP1, SPP2, SPP3) from PointPillar (PP)~\cite{lang2019pointpillars}, three sparse models (SCP1, SCP2, SCP3) from CenterPoint (CP)~\cite{yin2021center}, and one sparse model (SPN) from PillarNet (PN)~\cite{shi2022pillarnet}. These models quantized to use 8-bit multiplication and 32-bit accumulation, were assessed for computation savings and accuracy on well-known benchmarks like KITTI and nuScenes. Table~\ref{tab:cp_pn_perf} outlines the models, sparse convolution types, the average number of computations and sparsity, and model accuracy. The sparse models using SpConv (\eg~SPP1, SCP1) offered accuracy similar to dense baselines with lower computation savings. The models using SpConv-S (\eg~SPP3, SCP3), while significantly enhancing sparsity, suffered noticeable accuracy degradation. Conversely, the models with our SpConv-P (\eg~SPP2, SCP2) boosted computation savings without sacrificing accuracy, thus facilitating balanced sparsity-accuracy exploration in 3D object detection. Navigating these options necessitates diverse hardware support for various sparse convolution types, but it can lead to Pareto-optimal sparse convolution configurations, such as SPP2 and SCP2, within the spectrum of choices.

\subsection{Implementation Challenges}
\label{subsec:sp-net-challenges}

\noindent \textbf{Inadequate Sparse Architecture. }
\label{subsec:bad-sp-support}
Despite the computational benefits of sparse pillars, their efficient processing remains a challenge for current sparse Conv2D accelerators (SpConv2D-Acc)~\cite{parashar2017scnn, wang2021dstc, yang2021s2, shabani2023hirac}. The difficulty arises from the \textit{vector sparsity} of sparse pillars, characterized by sequences of zeros across all channels for each inactive pillar. This sparsity pattern differs from the element-wise activation sparsity commonly handled by SpConv2D-Acc where zeros appear at random locations as a result of the ReLU operation. 

Fig.~\ref{fig:latency-breakdown}(a) illustrates how SpConv2D-Acc processes sparse pillars. SpConv2D-Acc manages non-zero activation locations using sparse-compressed row (CSR) format~\cite{parashar2017scnn} or bitmasks~\cite{ wang2021dstc, yang2021s2, shabani2023hirac}. It transforms the convolution operation into matrix multiplication through \textit{im2col}, creating diagonal sparse patterns from the vector sparsity. SpConv2D-Acc treats these diagonal sparse patterns as individual sparse elements; it performs \textit{condensing} to compute matrix multiplication in output-stationary fashion. The resulting partial-sums are then scattered back into dense output buffers for further sparsification after full partial-sum accumulation and ReLU.

However, the vector sparsity of sparse pillars creates two problems in SpConv2D-Acc. First, the channel-wise sparsity of sparse pillars results in \textit{underutilization}, as entire rows of the processing element (PE) array remain idle because the condensed matrix does not exactly fit the size of the matrix unit. Second, each PE accumulates a vector of partial-sums corresponding to output pillars of different coordinates stored in the buffer, causing \textit{bank conflicts} (\eg~$O_{13}$ and $O_{23}$ access $Bank1$ at the same time in Fig.~\ref{fig:latency-breakdown}(a)). As sparsity increases, the output indices generated through the condensed matrix display an increasingly irregular pattern, leading to more severe bank conflicts.. Therefore, vector sparsity in highly sparse pillars cannot be effectively leveraged due to the combined impact of underutilization and bank conflicts. As Fig.~\ref{fig:latency-breakdown}(b) illustrates, these issues amplify as the computation sparsity increases.

\noindent \textbf{Inefficient Sparsity Support. }
\label{subsec:bad-sp-support}
While sparse convolution can theoretically lower the computational load in 3D object detection, its efficient implementation remains challenging~\cite{zhou2018voxelnet,yan2018second}. The major performance hurdle is the time-intensive process of mapping sparse input to output during convolution. Despite efforts by previous studies (\cite{tang2022torchsparse,yan2018second}) to enhance the efficiency of GPU-based sparse convolution algorithms, the mapping generation remains a bottleneck due to the limited parallelism and overhead of coordinate search. As shown in Fig.~\ref{fig:latency-breakdown}(c), Conv2D matrix multiplication takes up the majority of execution time in PP. In contrast, the total execution time for SPP variants does not decrease despite the decrease in convolution computation time due to the increased mapping overhead. These findings highlight the need for dedicated accelerators to handle sparse pillar convolution mapping more efficiently. While previous work has proposed cost-cutting strategies or designed custom sorting and cache units for faster searching~\cite{xu2019tigris,feng2020mesorasi,feng2022crescent}, none have specifically targeted sparse pillar convolution, suggesting substantial room for efficiency improvements in sparse pillar-based 3D object detection.

\noindent \textbf{Diverse Sparsity Patterns. }
\label{subsec:chal_diverse-sparsity}
Sparse convolution variants facilitate the study of accuracy-sparsity trade-offs (\S\ref{subsec:motivation}), yet the dynamic sparsity patterns and diverse convolution types necessitate adaptable hardware control. 
Fig.~\ref{fig:latency-breakdown}(d--f) illustrates the input-output pillar ratios (IOPR), which indicates the spatial expansion of pillars due to the dilation and the varying sparsity across sparse convolution layers. 
$BxCy$ labels the sparse convolution layers, where '$x$' represents the stage index, and '$y$' represents the convolution layer index within a stage, as shown in Fig~\ref{fig:fig2}(b). IOPR is a metric to identify the different vector sparsity patterns originated by different sparse convolutions. Initially, only a small number of active pillars are clustered together (Figure-1(c)), thus dilation after convolution (B1C1) results in the expansion of active pillars by almost two (i.e., IOPR=2). Standard SpConv in SPP1 gradually reduces sparsity due to dilation, and as the active pillar becomes denser, the degree of dilation decreases, and eventually, IOPR converges to 1 (Fig.~\ref{fig:latency-breakdown}(d)). Conversely, SpConv-S in SPP3 does not dilate pillars (i.e., IOPR=1), and thus the initial sparsity is maintained, but accuracy is sacrificed (Fig.~\ref{fig:latency-breakdown}(f)). We propose SpConv-P in SPP2 as a balanced solution, a dynamic pruning method for unimportant inactive pillars to maintain relatively high sparsity with minimal accuracy loss; IOPR increases periodically as dynamic pruning applied at the beginning of each stage spares more room for dilation due to its strided conv (Fig.~\ref{fig:latency-breakdown}(e)). This diverse nature of sparsity patterns depending on the sparse convolution type necessitates a method to change dataflow adaptively depending on the observed sparsity for high hardware utilization.



\begin{figure}[t]
\begin{center}
\centerline{\includegraphics[width=0.97\linewidth]{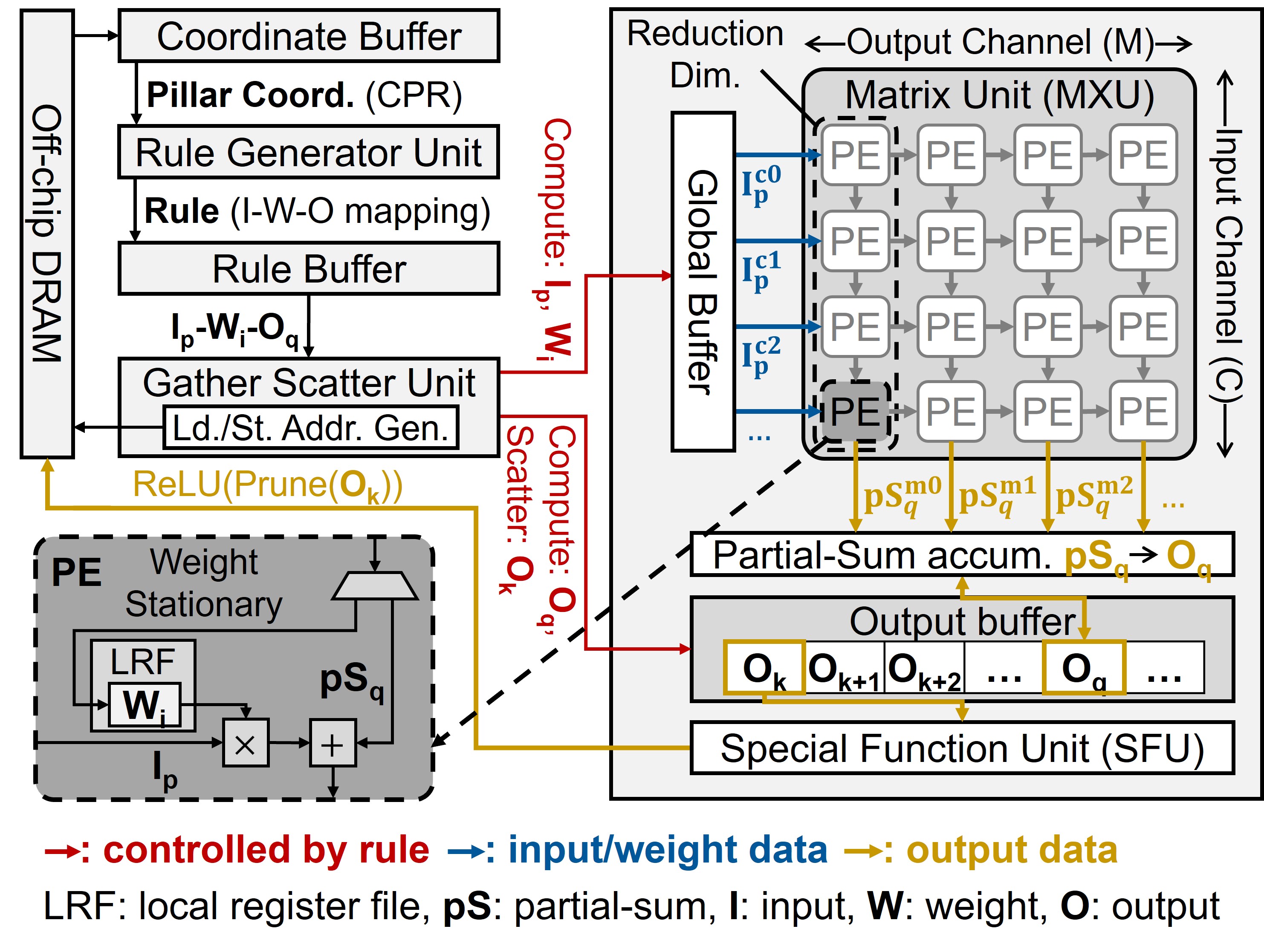}}
\caption{Overall architecture of \arch.}
\label{fig:arch_overview}
\end{center}
\vspace{-0.6cm}
\end{figure}

\section{Architecture}
\label{sec:arch}

This section proposes \arch, novel \textit{vector sparsity} support for accelerating sparse pillar convolution. This method eliminates the need for inefficient coordinate searching for input-output mapping. The key insight is that if the pillar coordinates are once orderly aligned, their ordering can be maintained throughout sparse convolution to avoid global search and generate mapping in a constant time. In this section, we discuss the proposed mapping generation algorithm and supporting hardware architectures that realize its expected efficiency.

\begin{figure*}[t]
\begin{center}
\centerline{\includegraphics[width=\linewidth]{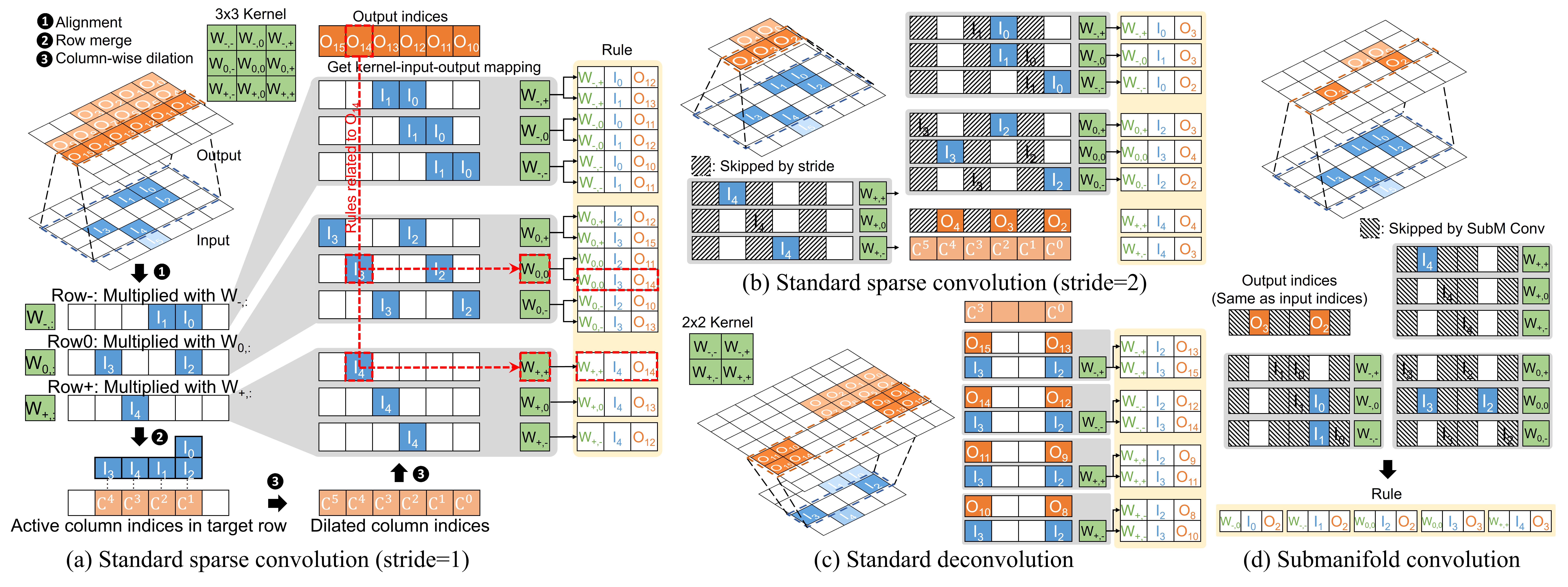}}
\caption{Rule generation algorithm of \arch.}
\label{fig:spconv_overview}
\end{center}
\vspace{-0.6cm}
\end{figure*}

\begin{figure*}[t]
\begin{center}
\centerline{\includegraphics[width=2\columnwidth]{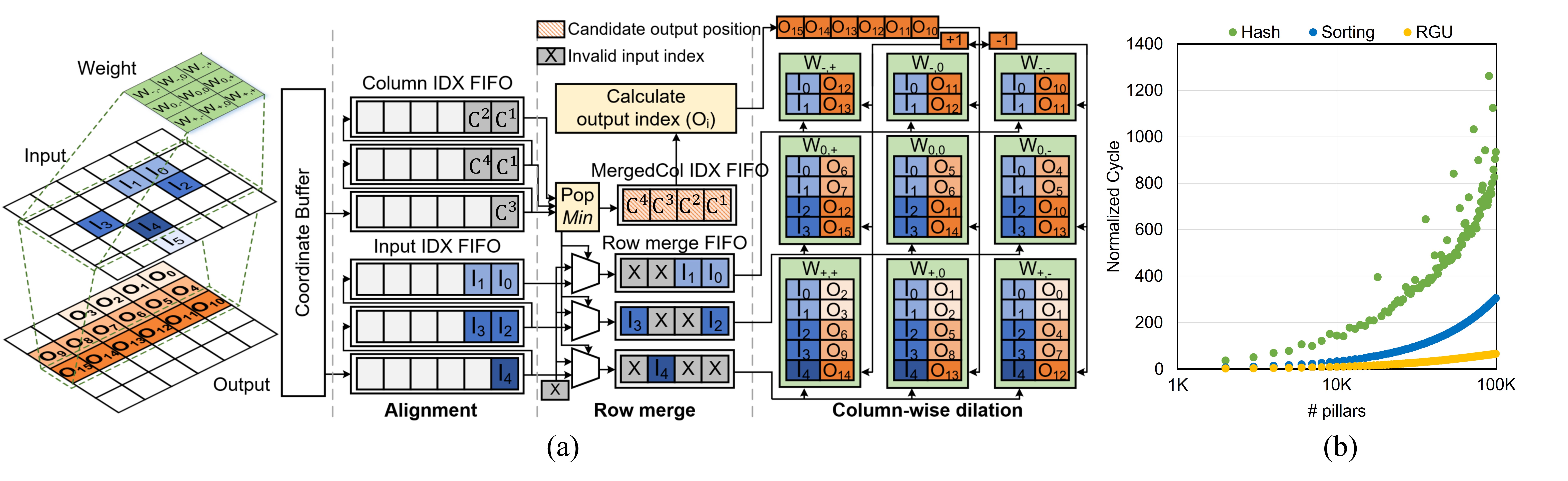}}
\caption{(a) Hardware architecture of rule generator unit (RGU). C$_{*}$, I$_{*}$, W$_{*}$, and O$_{*}$ denote index of column, input, kernel, and output, respectively. $\times$ denotes an invalid signal. (b) Comparison of rule generation methods: hash table, sorting, RGU (ours).}
\label{fig:rule-gen_archi}
\end{center}
\vspace{-0.9cm}
\end{figure*}

\subsection{Overview}
\label{subsec:arch_overview}

The overview of \arch architecture is shown in Fig.~\ref{fig:arch_overview}. The Rule Generation Unit (RGU) uses the coordinate data of active pillars, formatted as compressed-pillar-row (CPR), to efficiently create an input-output index mapping, referred to as a ``rule", for the corresponding sparse pillar convolution. Leveraging this explicit mapping information, the Gather-Scatter Unit (GSU) can retrieve \textit{only} active input pillars to the global buffer and place partial-sums in the output buffer at the respective addresses calculated by this unit, enabling sparse pillar convolution to be managed like simple matrix multiplication via the Matrix Unit (MXU). Furthermore, the GSU includes a Special Function Unit (SFU) to accumulate partial-sums and prune pillars for on-the-fly sparsification.  

\arch distinguishes itself from SpConv2D-Acc with key architectural differences. Unlike SpConv2D-Acc, \arch processes sparse convolution in a \textit{weight-stationary} manner, enabling the complete channel elements of an active pillar to be fetched together across rows, thus \textit{avoiding underutilization}.  Furthermore, a vector of partial-sums across the output channel produced by the MXU corresponds to a \textit{single} pillar coordinate and can be updated into the single-bank output buffer \textit{without conflict}. 
This seamless mapping of vector sparsity into a 2D systolic array is possible thanks to RGU's explicit coordinate mapping of (I-W-O) and the GSU's efficient management of sparse data. Subsequent sections detail how the RGU and GSU provide effective support for a variety of sparse convolutions.

\subsection{Rule Generation Unit}
\label{subsec:arch_rule-gen}

The goal of RGU is to create the mapping information (called a rule) that indicates which inputs and weights are convolved to produce an output pillar. Since the active input pillars are sparely located, a different set of inputs is involved in producing each output pixel. Finding a rule usually requires a cumbersome neighbor search, but we can exploit the limited window size of sparse convolution to simplify the rule generation (RuleGen). 

\subsubsection{Algorithm}
\label{subsec:arch_rule-gen-algo}

The RuleGen algorithm consists of three steps, \textit{alignment, row merge, and column-wise dilation}, to generate the mapping of ($I_p$, $W_i$, $O_q$) to indicate which weight index ($i$) and input index ($p$) are involved to compute the output with an index $q$. Fig.~\ref{fig:spconv_overview}(a) illustrates these steps for SpConv with 3x3 weights.

\noindent \textbf{Alignment (\circled{1}).} The pillar data encoded in CPR (a sparse row-wise encoding of pillar coordinates similar to compressed sparse row) is initially fetched as input. As an example, the three rows of active inputs $I_0,I_1,I_2,I_3,I_4$ are aligned by column index to generate the rule for the third output row. Since the active inputs are encoded by CPR, their indices monotonically increase within each row. Note that the three rows of input are associated with the top, center, and bottom ($W_{-,:},W_{0,:},W_{+,:}$) rows of weight.

\noindent \textbf{Row Merge (\circled{2}).} After the alignment, each aligned input row is associated with the row of weight while the column indices of the active inputs can be merged across the rows, \eg~($C^1$,$C^2$) + ($C^1$,$C^4$) + ($C^3$) = ($C^1,C^2,C^3,C^4$) where $x$ in $C^x$ denotes a column index. 

\noindent\textbf{Column-wise Dilation (\circled{3}).} Then, each of these merged columns is dilated horizontally by +/-1 step  (middle column of Fig.~\ref{fig:spconv_overview}(a)) along with the input and weight indices to find the mapping for the active output ($O_{10},O_{11},O_{12},O_{13},O_{14},O_{15}$). 

Thanks to CPR, the output indices also monotonically increase. Each mapping of input-weight-output indicates the index needed for convolution computation associated with each weight: e.g., $O_{14}=I_{3}\cdot W_{0,0} + I_{4}\cdot W_{+,+}$. This mapping information is stored in the rule buffer organized by weight for weight-stationary sparse convolution. 

The rules for other sparse convolution operations, as shown in Fig.~\ref{fig:spconv_overview} (b--d), can be generated similarly to SpConv. Sparse strided convolution (SpStConv) follows the same RuleGen procedure as SpConv, but omits odd-numbered columns from the output when the stride is 2. Sparse deconvolution (SpDeconv) differs in that RuleGen expands the input row rather than merging it, and there's no overlap in output coordinates due to non-overlapping receptive fields. In SpConv-S, output location is limited to active input locations, ignoring the dilation effect. For SpConv-P, the RuleGen procedure mirrors SpConv, but the output pillars are pruned to serve as active input pillars for the next SpConv. 

Note that the complexity of the proposed RuleGen is $O(P)$ where $P$ is the number of active pillars, thanks to the straightforward three-step search for mapping input and output. Thus, RGU can swiftly provide necessary mapping information for sparse convolution operations.

\subsubsection{Hardware Architecture}
\label{subsec:arch_rule-gen-hw}

The proposed RGU is a streaming hardware architecture designed for the RuleGen algorithm. RGU consists of the same three stages as its algorithm: alignment, row merge, and column-wise dilation (Fig.~\ref{fig:rule-gen_archi}(a)). In the alignment stage, active input indices (\eg~$I_0,I_1|I_2,I_3|I_4$) and corresponding column coordinates (\eg~$C^1,C^2|C^1,C^4|C^3$) are aligned row-by-row via a FIFO chain (see the second column of Fig.~\ref{fig:rule-gen_archi}(a)). The FIFO chain cycles through the input row, with each row passing its contents upwards and accepting new input at the bottom, enabling structured input processing in the row merge stage. 

\begin{figure*}[t]
\begin{center}
\centerline{\includegraphics[width=0.97\linewidth]{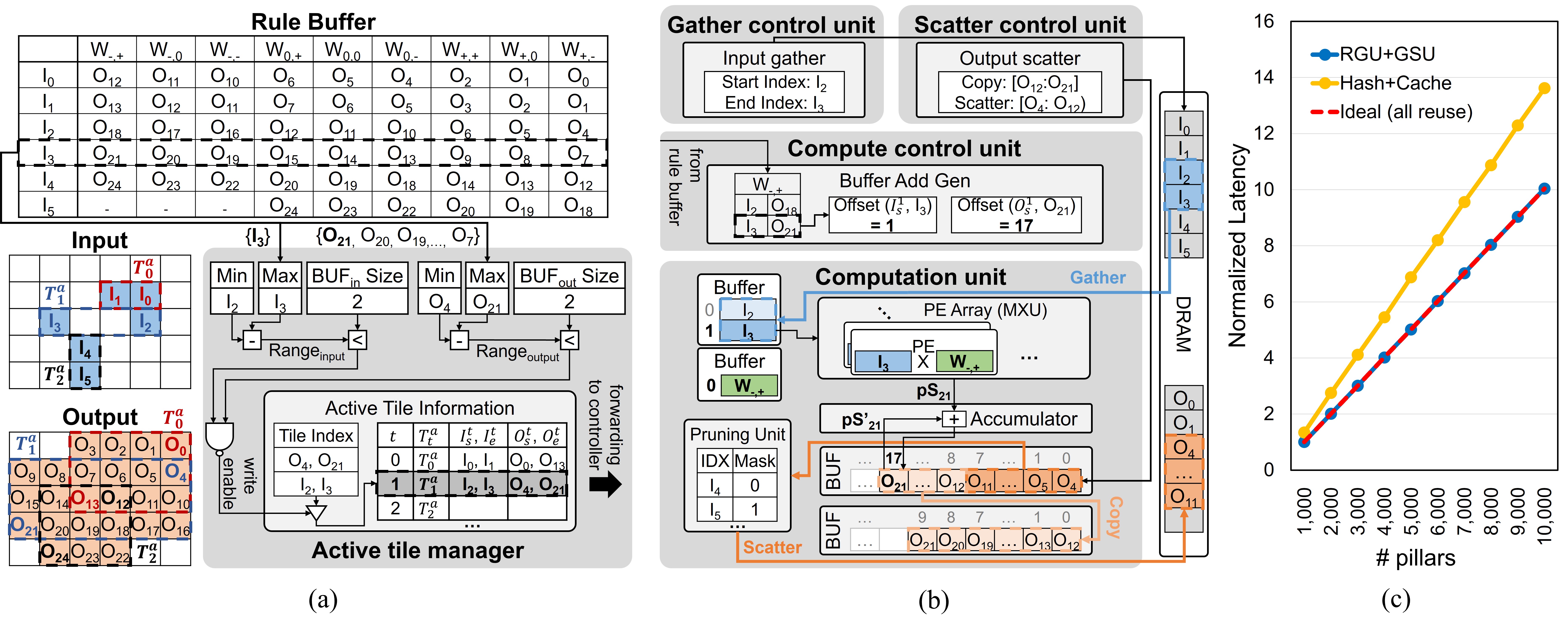}}
\caption{Gather-Scatter Unit (GSU): (a) active tile management algorithm, (b) GSU architecture, (c) DRAM latency comparison. }
\label{fig:g_s_archi}
\end{center}
\vspace{-0.6cm}
\end{figure*}

In the row merge stage, the active input pillars for the smallest column index (e.g., $I_0$ and $I_2$ for $C^1$) are popped from the chained three input index FIFOs and sent to the row merge FIFOs,  where each entry corresponds to the active input pillars associated with a particular column index. The vertical position of the input pillars among the top, center, and bottom is associated with the weight row of the kernel for vertical dilation (e.g., $I_0$ and $I_2$ correspond to $W_{-,:}$ and $W_{0,:}$, respectively). Finally, in the column-wise dilation stage, the output column indices (e.g., $O_{10},O_{11},...,O_{15}$) are calculated using the merged columns $C^1,C^2,C^3,C^4$, and each of the row merge FIFOs feeds three rule buffers according to their horizontal dilation. For example, $I_{4}$ results in mapping: $W_{+,-}$ with $O_{12}$, $W_{+,0}$ with $O_{13}$, and $W_{+,+}$ with $O_{14}$.

Each stage can be pipelined, calculating necessary information without trial and error, thus generating rule information every cycle. Notably, within each rule buffer, input and output indices are automatically sorted in ascending order due to the CPR-based streaming process of input, significantly simplifying dataflow management.

\subsubsection{Analysis}
\label{subsec:arch_rule-gen-comp}

The primary purpose of RuleGen is to identify unique output indices from all potential combinations of active inputs and weight kernels, assisting in partial-sum accumulation. To evaluate the effectiveness of the RGU, we contrast it with two prevalent RuleGen methods: 
the hash table-based approach widely used in GPU implementation, Spconv-Libary~\cite{spconv2022}, and the sorting-based approach used in the state-of-the-art point cloud accelerator, PointAcc~\cite{lin2021pointacc}.
Hash tables are common data structures for querying input-output index mappings. Although theoretically constant in query time, hash tables require additional cycles and hardware resources to parallelize processing and resolve collisions through chaining. To improve this, merge sorter-based RuleGen implemented was proposed in \cite{lin2021pointacc}. By sorting all active input positions based on kernel offsets, it identifies unique output coordinates via an intersection map and utilize a bitonic merger to reduce sorting complexity from $O(P\times log(P))$ to $O(log(N)\times log(P/N)\times (P/N)$ where $P$ is the number of pillars and $N$ is the length of bitonic merger. In contrast, RGU requires only $O(P)$ time and hardware resources, attributed to its streaming process of RuleGen.

To quantitatively compare RGU with the hash table and merge sorting methods, we configured a hash table-based RuleGen following the basic settings of Spconv-Libary; the main table size is set as $2 \times P$ with additional memory to hold up to $K \times P$ chains to resolve collisions ($K=9$ for a $3 \times 3$ kernel window). Also, we implemented an $N$-length ($N=64$) bitonic merger for the merge sorter for the sorting-based RuleGen, matching the design spec of PointAcc. Fig.~\ref{fig:rule-gen_archi}(b) shows the normalized mapping cycle for the hash table, merge sorter, and RGU methods in relation to the number of active pillars. Our experiments involving up to 100,000 pillars showed that RGU is, on average, $5.9\times$ faster than the hash table due to significant overheads in collision resolution for multiple input coordinates contributing to the common output coordinates. Furthermore, RGU is $3.7\times$ faster than the merge sorter, thanks to its lower processing complexity for identifying input-output mapping. These results substantiate the efficiency of the proposed RGU method.

\begin{figure*}[]
\begin{center}
\centerline{\includegraphics[width=1.97\columnwidth]{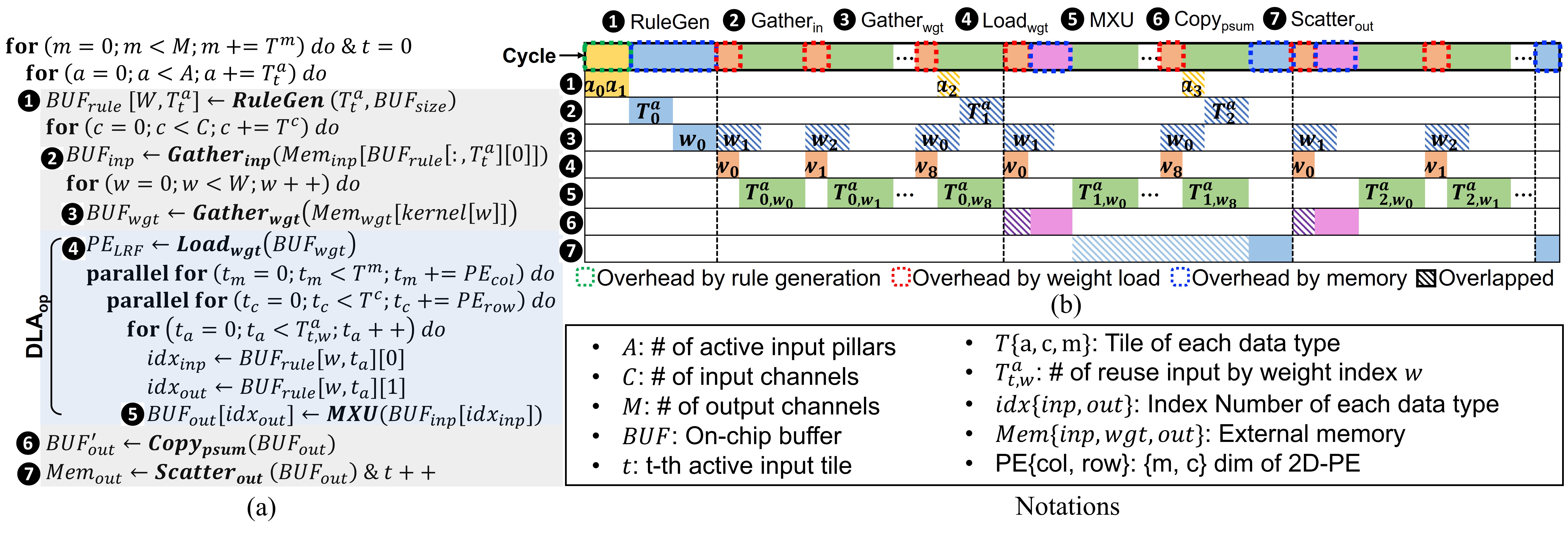}}
\caption{\arch dataflow: (a) pseudo-code, (b) timing diagram.}
\label{fig:dataflow}
\end{center}
\vspace{-0.7cm}
\end{figure*}

\subsection{Gather-Scatter Unit}
\label{subsec:arch_g-s}

The gather-scatter unit (GSU) orchestrates the data collection and distribution of inputs and outputs using rule information from RGU to control MXU execution. This section explains how the GSU handles sparse inputs and outputs and manages \arch execution.  

\subsubsection{Active Tile Management}
\label{subsec:arch_g-s-algo}

While RGU provides the mapping rule between input and output pillars for SpConv, the data locations of active input and output can be arbitrary. Consequently, when conducting SpConv computation based on either output or input order, irregular memory accesses are inevitable for gathering input or scattering output. As a result, numerous point-cloud accelerators, such as PointAcc, employ expensive cache mechanisms for data reuse \cite{lin2021pointacc}. However, even with these mechanisms, they still encounter problematic cache misses.

To overcome this issue, \arch introduces an active tile manager (ATM) that leverages the consistent progression of input-output indices in CPR format for efficient dataflow management. Fig.~\ref{fig:g_s_archi}(a) shows the proposed active tile management algorithm. Active tile size is constrained by the on-chip buffer size: the number of active inputs or outputs must not exceed the buffer's capacity. Given the monotonic increase of input and output indices across all rule buffers, we can define start ($I^t_s/O^t_s$) and end ($I^t_e/O^t_e$) indices, along with tile size ($|BUF_{in/out}|$) for loading input/output. Thanks to the continuous nature of indices, loaded input and output are guaranteed to be reused fully throughout the convolution computation. This active tile information can be determined by traversing the rule buffer only once.

\subsubsection{Sparse Convolution Execution}
\label{subsec:arch_g-s-algo}
ATM streamlines sparse convolution execution on \arch, as shown in Fig.~\ref{fig:g_s_archi}(a) and (b). The rule generated by RGU is filtered by ATM to construct active tile information ($I^t_s=I_2$ and $I^t_e=I_3$), which is relayed to the gather and scatter control units. Under the direction of the compute control unit , the input data and pre-loaded weight ($W_{-,+}$) staged in the input buffer ($BUF_{in})$ initiate execution of the computation unit. The partial-sums ($pS_{21}$) calculated by the 2D PE array are accumulated and then scattered into the output buffer ($BUF_{out}$) via offset address ($17$) until fully accumulated. The addresses of the data used for operations related to a specific rule are directly calculated by determining the offset value between the index of corresponding data and the start index received from the ATM. Fig.~\ref{fig:g_s_archi}(b) also illustrates two additional cases: 1) when outputs overlap for consecutive input tiles, $pS$'s from one output buffer are copied to another for further accumulation, and 2) during SpConv-P processing, the fully accumulated output is passed through the pruning unit to prune inactive pillars.

\subsubsection{Analysis}
\label{subsec:arch_g-s-algo}

We evaluated the efficiency of ATM by comparing its data access performance with cache-based sparse dataflow management. Fig.~\ref{fig:g_s_archi}(c) illustrates the normalized DRAM latency (assessed on Ramulator) of both methods as the number of active pillars increases, alongside an ideal theoretical DRAM latency. The cache-based method employs a hash table for RuleGen and a direct-map cache, standard elements of typical SpConv implementations~\cite{graham2015sparse,yan2018second,tang2022torchsparse}. We set the cache and $BUF_{in}$ sizes to 32KByte, managed the hash table in an input-stationary manner to optimize input reuse (\cf~\cite{tang2022torchsparse}), and set the cache block size to $64$ to minimize cache misses, as suggested by \cite{lin2021pointacc}. As depicted in the figure, our sparse dataflow management (RGU+GSU) outperforms the cache-based method (Hash+Cache), with the performance gap widening as the number of active pillars increases. The trace analysis revealed that the cache-based method suffered multiple input fetches, particularly near active output tile boundaries. Conversely, by exploiting the monotonicity of input-output indices, our simple hardware setup ensures full data reuse and expedites sparse convolution calculations, matching the latency of the all-reuse DRAM accesses.

\subsection{\arch Dataflow}
\label{subsec:arch_g-s-arch}

\subsubsection{Configurable Dataflow}

The RGU and GSU microarchitectures efficiently manage sparse input and output data, enabling the MXU to handle sparse pillar convolution as tiled matrix multiplication. We further developed control units for RGU and GSU to configure the \arch dataflow for various types of sparse convolutions. Provided a series of sparse convolution computations tiled along the channels (input: $C$, output $M$) and active pillars ($A$), the \arch dataflow can be defined with the subsequent instructions: (1) $RuleGen$: generate a rule for a tile of active pillars; (2) $Gather_{inp}$: Collect a tile of sparse input data; (3) $Gather_{wgt}$: Collect a tile of weight data, (4) $Load_{wgt}$: Load weight date into local register file of PE; (5) $MXU$: Execute PE-array for matrix multiplication; (6) $Copy_{psum}$: Copy boundary partial-sums; and (7) $Scatter_{out}$: store final output to memory.

\begin{figure*}[]
\begin{center}
\centerline{
\includegraphics[width=0.97\linewidth]{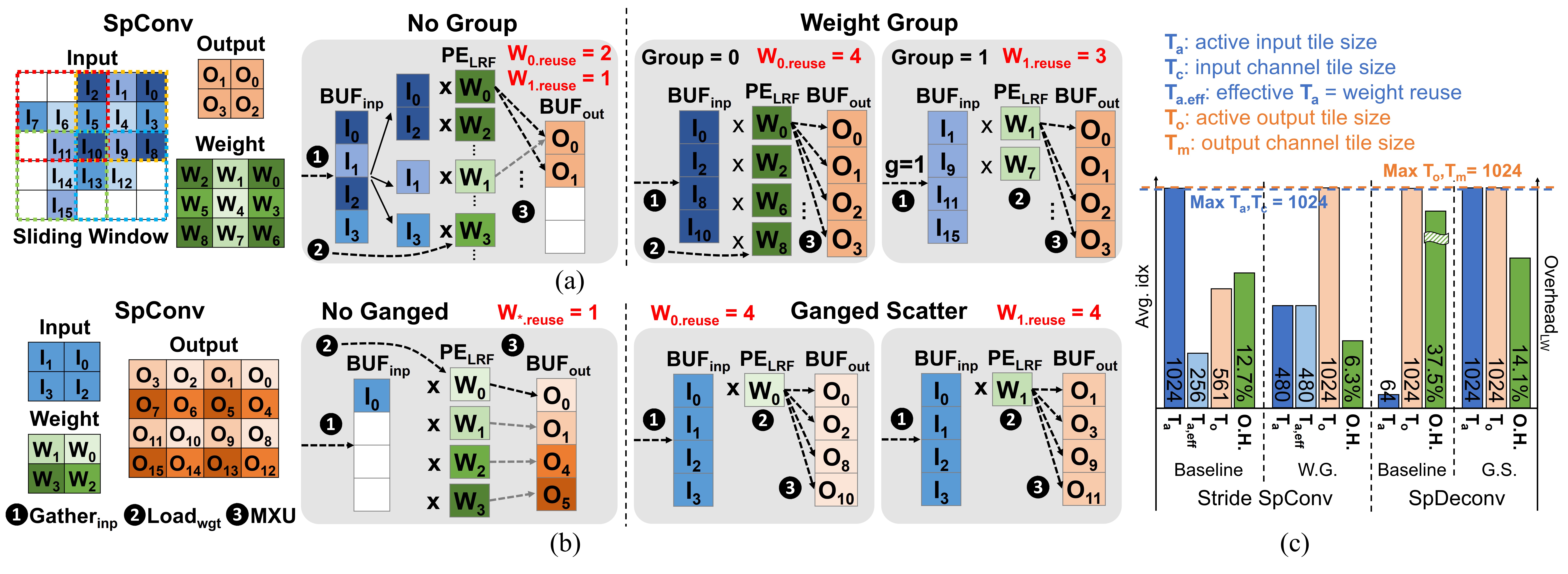}
}
\caption{Dataflow optimization techniques: (a) weight grouping for SpStConv, (b) ganged scatter for SpDeconv. (c) Impact of dataflow optimization on overhead reduction ($T_{a\_eff}$: effectively reused active inputs).
}
\label{fig:dataflow_opt}
\end{center}
\vspace{-0.5cm}
\end{figure*}

The seven instructions in \arch are designed to set up its dataflow for sparse convolution: $RuleGen$ for input-output mapping, ${Gather}_{in}$, ${Gather}_{wgt}$, and ${Scatter}_{out}$ for collecting data from the external memory or scattering it out, ${Load}_{wgt}$ for loading weights into the local register file, and ${Copy}_{psum}$ for output buffer management. Fig.~\ref{fig:dataflow}(a) illustrates the pseudocode defined by these instructions. The inner computation (marked in blue) is performed in a weight-stationary manner as channel dimensions ($T_{m}$ and $T_{c}$), determined offline from the layer specification, can be spatially assigned to the PE array. Conversely, the outer computation (marked in gray) operates in an output-stationary manner since the partial-sum reduction is typically costlier than input due to higher accumulation precision and the necessity for double read/write access for output scatter. Unlike previous works (\eg~\cite{lin2021pointacc}), \arch can be tailored for variable tile sizes $T^m, T^c, T^a$ using the proposed seven instructions, proving particularly beneficial for performance optimization.

Fig.~\ref{fig:dataflow}(b) presents a sample timing diagram. It is important to note that the execution cycles of certain instructions cannot overlap with MXU computation cycles. For example, $Load_{wgt}$ cycles cannot be hidden, requiring stalls in the PE array for weight copying to local register files. Similarly, $Copy_{psum}$ halts the pipelined computation for partial-sum accumulation across the consecutive active input tiles since the partial-sums must be copied before starting a new computation set. On the contrary, instructions like $RuleGen$, $Gather_{inp}$, $Gather_{wgt}$, and $Scatter_{out}$ can hide their cycles after the first run, given they're double-buffered. However, in certain scenarios, $Scatter_{out}$ cycles may spill over due to an insufficient number of MXU cycles for a smaller $T^a$. Thus, proper configuration of tile sizes is essential to minimize non-overlapping cycles.


\subsubsection{Dataflow Optimization Techniques}

Instructions like $Load_{wgt}$ introduce non-overlapping overhead, significantly affecting \arch's execution time. To mitigate this overhead, it is crucial to maximize weight reuse by increasing the number of input data sharing the same weight, referred to as the active pillar tile $T^a$. The GSU is capable of adaptively adjusting this tile size to accommodate various sparsity patterns as described in Fig.~\ref{fig:latency-breakdown}(d-f). This adaptive strategy works particularly effectively for the standard sparse convolution operations (SpConv, SpConv-S, SpConv-P), but its impact is limited for SpStConv and SpDeconv. In this section, we will introduce unique dataflow optimization techniques designed to efficiently handle all types of sparse convolution operations.

\noindent \textbf{Weight Grouping. }
In SpStConv, a fixed stride in convolution operations restricts input reuse. But a closer analysis reveals new opportunities: weights that follow the stride pattern can reuse the input. Notably, in the case of stride=2, weight grouping of \{0,2,6,8\}, \{1,7\}, \{3,5\}, and \{4\} shares the strided input. For instance, with the default dataflow, the sequentially gathered active inputs are only used once or twice by the weights (Fig.~\ref{fig:dataflow_opt}(a)-left). Whereas, if the active inputs are gathered according to the weight group \{0,2,6,8\}, the corresponding inputs $I_{0}$, $I_{2}$, $I_{8}$, $I_{10}$ will be fully used by the weights (Fig.~\ref{fig:dataflow_opt}(a)-right). In Fig.~\ref{fig:dataflow_opt}(c), the left figure illustrates the performance improvement achieved through weight grouping in SpStConv, using the first SpStConv of SPP2 as an example. Without weight grouping, there is a large number of active pillars ($T_a$), but only a small subset ($T_{a,eff}$) is effectively used due to stride, leading to increased weight load overhead. However, with weight grouping, we can generate a significantly higher number of outputs ($T_o$) thanks to the stride-matched weight ordering. This, in turn, increases the number of active pillars ($T_{a,eff}$) that can reuse the weights, reducing the overall overhead. When applied to the first SpStConv layer of SPP2, this reduction cuts the existing 12.7\% overhead down to 6.3\%, as demonstrated in the figure.

\begin{figure*}[h]
\begin{center}
\centerline{
\includegraphics[width=\linewidth]{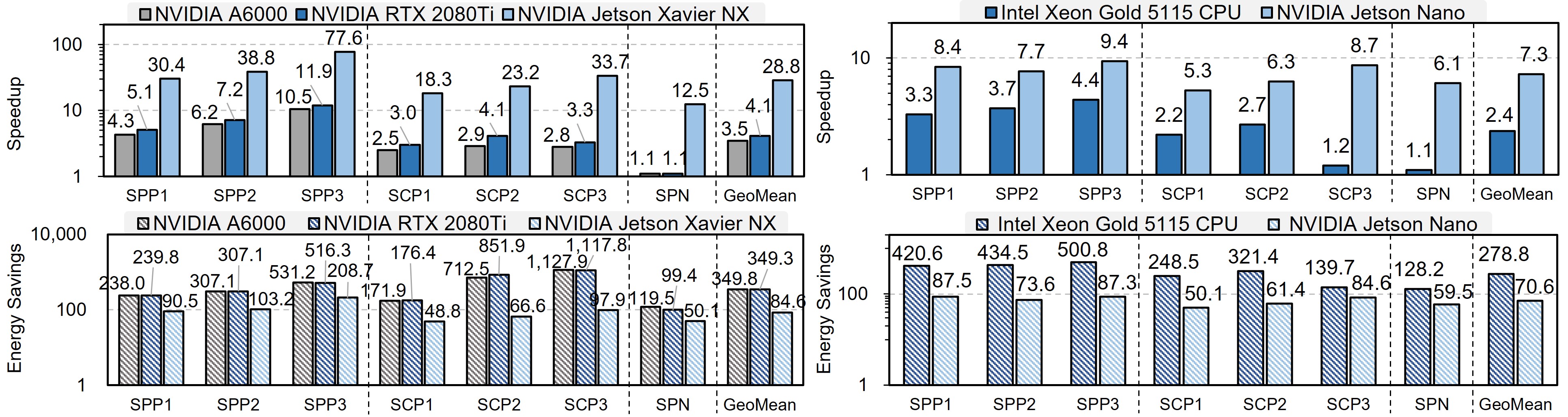}
}
\caption{Speedup and energy savings of \arch (left: high-end, right: low-end) for sparse pillar-based 3D objection: PointPillars (SPP1--3), CenterPoint (SCP1--3), and PillarNet (SPN).}
\label{fig:sparse_speedup}
\end{center}
\vspace{-0.7cm}
\end{figure*}

\noindent \textbf{Ganged Scatter. }
For SpDeconv, accumulation is not required as each input directly corresponds to specific outputs. Consequently, the default output-stationary dataflow proves inefficient during the scatter-out phase due to lack of reuse. Instead, we propose \textit{ganged scatter} to group output production for the same weight. We adjust \arch's dataflow to scatter out a group of output immediately after the computation with each loaded weight. Fig.~\ref{fig:dataflow_opt}(b) provides an illustration; in the default dataflow, the scatter is performed after computing all computations for all kernels, resulting in fewer active pillars to sacrifice weight reuse. However, by performing scatter for each kernel separately, output pillars can be generated for an equal number of active pillars, increasing weight reuse. This allows us to gather more active pillars. As shown in the right figure of Fig.~\ref{fig:dataflow_opt}(c), in the case of the third SpDeconv layer of SPP2, a 16-fold increase in weight reuse reducing the overhead from 37.5\% to 14.1\%.


\section{Evaluation}
\label{sec:eval}

\subsection{Evaluation Setup}

\noindent \textbf{Benchmarks. } 
To evaluate \arch's performance, we utilized three state-of-the-art 3D object detection networks based on point pillars: PP, CP, and PN. We conducted diverse evaluations using the KITTI dataset for PP and the nuScenes dataset for CP and PN. We created sparse models for each network that incorporate three types of convolutions: SpConv, SpConv-P, and SpConv-S. Detailed model information is provided in Table~\ref{tab:cp_pn_perf}. These extensive sparsity-conscious benchmarks allow for a thorough appraisal of \arch's performance in relation to speed-sparsity trade-offs.

\noindent \textbf{Hardware Implementation. } 
To evaluate the proposed accelerator architecture's efficiency, we implement \arch in Verilog and verify its design through RTL simulations. Synopsys Design Compiler synthesizes \arch at 1GHz clock frequency under SAED 32nm technology. We simulate \arch's power consumption using annotated switching activity from selected benchmarks. We create a cycle-accurate simulator to model hardware behavior, calculate cycle counts, and read/write on-chip SRAMs and an off-chip DRAM. The simulator, verified against the Verilog implementation, is integrated with Ramulator~\cite{kim2015ramulator} to model DRAM behaviors. We obtain SRAM energy using CACTI \cite{mura2015cacti} and DRAM energy from Ramulator-dumped DRAM command traces. For comparison, \arch is configured into high-end (HE) and low-end (LE) types; HE and LE employ a $64\times64$ (8 TOPS) and $16\times16$ (512 GOPS) systolic array for MXU, respectively.

\noindent \textbf{Baseline. } 
For evaluation baselines, we used NVIDIA A6000 and RTX 2080Ti (GPUs) and NVIDIA Jetson Xavier NX (Jetson-NX) for HE, and Intel Xeon 5115 (CPU) and NVIDIA Jetson Nano (Jetson-NN) for LE (NVML and Turbostat for GPU and CPU power measurements). For pillar-based 3D object detection, we implemented PP and CP using Conv2D in PyTorch (matrix operations with Intel MKLDNN / cuDNN) \cite{cudnn}, while SPP, SCP, PN, and SPN were implemented with the SpConv library~\cite{yan2018second}, which was the only open-source library currently supporting all variants of sparse convolution, including SpDeconv. For a direct ASIC-level comparison, we also developed DenseAcc (both HE and LE), a simplified version of \arch that supports only dense convolution operations without RGU, GSU, and pruning support.

\begin{figure*}[h]
\begin{center}
\centerline{
\includegraphics[width=0.97\linewidth]{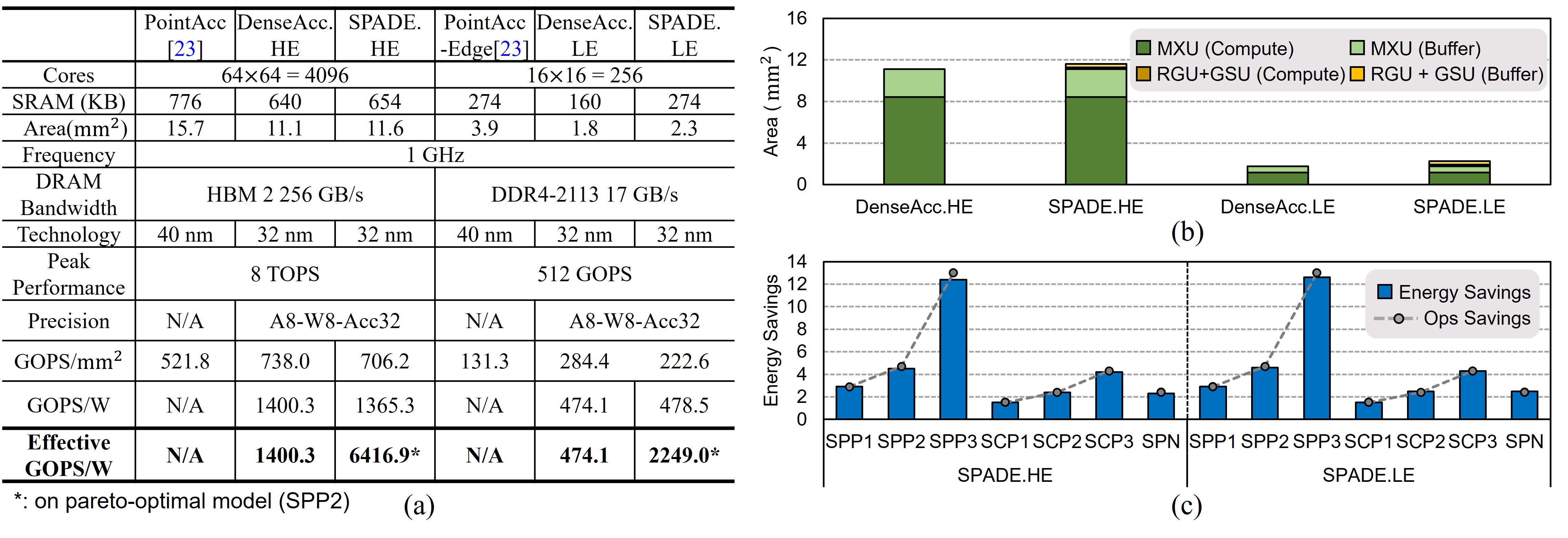}
}
\caption{Hardware evaluation: (a) accelerator comparison, (b) area breakdown, (c) energy savings over ideal dense accelerators.}
\label{fig:table2_comp_area_speedup}
\end{center}
\vspace{-0.6cm}
\end{figure*}

\subsection{Evaluation Results}
\label{sec:eval_results}

\begin{figure*}[t]
\begin{center}
\centerline{
\includegraphics[width=0.97\linewidth]{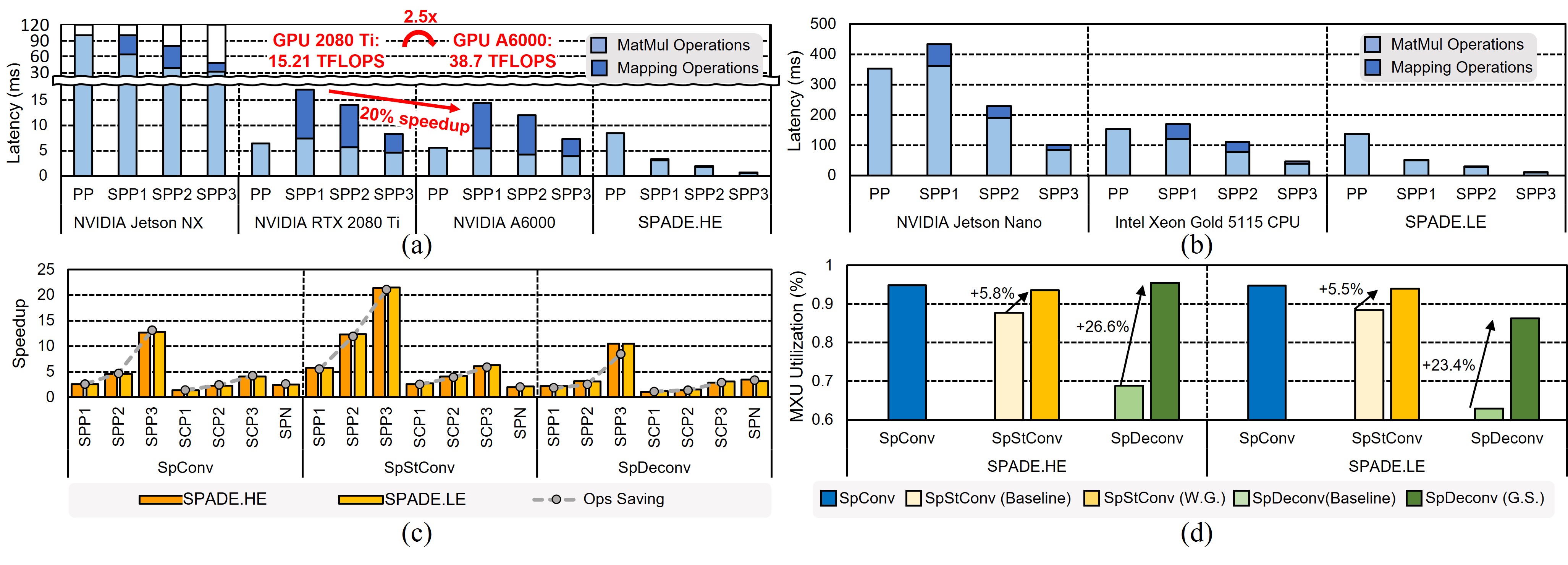}
}
\caption{Latency comparison of \arch with server and edge platforms: (a) high-end, (b) low-end. (c) Speedup breakdown on various sparse convolutions. (d) Improvement of utilization from dataflow optimization.}
\label{fig:lat_break_sparse}
\end{center}
\vspace{-0.6cm}
\end{figure*}

\subsubsection{Speedup and Energy Savings}

Fig.~\ref{fig:sparse_speedup} shows the speedup and energy savings of \arch compared to GPUs and Jetson-NX for HE and CPU and Jetson-NN for LE, evaluated on the sparse models. On average, \arch.HE achieves $3.5\times$, $4.1\times$ and $28.8\times$ speedup, and $349.8$, $349.3\times$ and $84.6\times$ energy savings over A6000, 2080 Ti and Jetson-NX, respectively. As sparsity increases, both speedup and energy savings generally rise, peaking at the highest sparsity for each sparse model. However, SPN performance is relatively low due to the already sparse baseline (PN). A similar trend is observed for LE, except for the exceptional CPU performance on SCP3. While GPU and CPU outperform Jetson-NX and Jetson-NN in speedup, their energy efficiency is lower. \arch demonstrates speedup with a range of 1.1$\times$ to 77.6$\times$ and energy savings with 48.8$\times$ to 1,117.8$\times$ for HE.

\subsubsection{Hardware Evaluation}

Fig.~\ref{fig:table2_comp_area_speedup}(a) presents a comparison of \arch's hardware performance to DenseAcc and PointAcc\cite{lin2021pointacc} for both HE and LE. When matched in form factors (including cores, frequency, on-chip memory, technology, and DRAM bandwidth), \arch uses smaller area and SRAM size compared to PointAcc. Also, \arch offers competitive performance density compared to DenseAcc; Despite incorporating sparsity support, \arch exhibits only slightly lower peak $GOPS/mm^2$ and $GOPS/W$. Furthermore, \arch achieves significantly higher \textit{effective} $GOPS/W$ when running a sparse model. For SPADE.HE and LE running SPP2 (sparsity=73.5\%), the effective $GOPS/W$ increases by 4.6$\times$ and 4.7$\times$, respectively. This improvement stems from \arch's ability to leverage sparsity to skip meaningless computations, thanks to its hardware support for sparsity, while DenseAcc processes dense convolution by densifying sparse pillars.

Fig.~\ref{fig:table2_comp_area_speedup}(b) displays the area breakdown for DenseAcc and \arch in HE and LE configurations. In \arch.HE, additional hardware components occupy only 4.3\% of the total area. Although the extra hardware portion is larger in \arch.LE, the total area of \arch remains significantly smaller than PointAcc. In addition, Fig.~\ref{fig:table2_comp_area_speedup}(c) measures energy savings, with \arch achieving near-optimal scaling in savings compared to DenseAcc running on PP. These results emphasize the efficiency of the proposed RGU and GSU.

\subsubsection{Source of Performance Gain}

\noindent \textbf{Rule Generation Unit. }
Fig.~\ref{fig:lat_break_sparse}(a) and (b) display the latency breakdown of dense and sparse PointPillars on various HE and LE platforms (only SPP1-3 are shown, but the trends are consistent across all cases). GPUs, CPU, and Jetson platforms rely on Spconv-Library~\cite{spconv2022}, which uses a hash table and cache for sparse execution. These platforms experience slowdowns due to inefficient mapping, while \arch exhibits exceptional efficiency in rule generation (\ie, minimal time spent on mapping) with the proposed RGU. In particular, the performance comparison between A6000 and 2080Ti underscores GPU inefficiencies (A6000 offers 2.5x peak throughput over the 2080Ti but only achieves a 20\% speedup), emphasizing the need for customized acceleration.

\noindent \textbf{Efficiency on Diverse Sparsity Operations. }
Fig.~\ref{fig:lat_break_sparse}(c) demonstrates the OPs savings and speedup realized by \arch when compared to DenseAcc for HE and LE, respectively, across different types of sparse convolutions used in each model. \arch achieves a speedup that aligns directly with OPs savings, showcasing the maximum potential speedup across various sparse convolutions, thanks to the GSU's ability to capture sparsity effectively.

Fig.~\ref{fig:lat_break_sparse}(d) displays MXU utilization, further confirming \arch's improved efficiency in handling diverse sparse convolutions. SpConv is mostly efficient, achieving over 90\% MXU utilization consistently. In contrast, SpStConv and SpDeconv experience a poor MXU utilization ($<$70\%) in the baselines due to inefficient sparsity patterns, but the proposed dataflow optimization techniques, weight grouping, and ganged scatter notably reduce overhead, reaching nearly 90\% utilization. 

\begin{figure}[t]
\begin{center}
\centerline{
\includegraphics[width=0.87\linewidth]{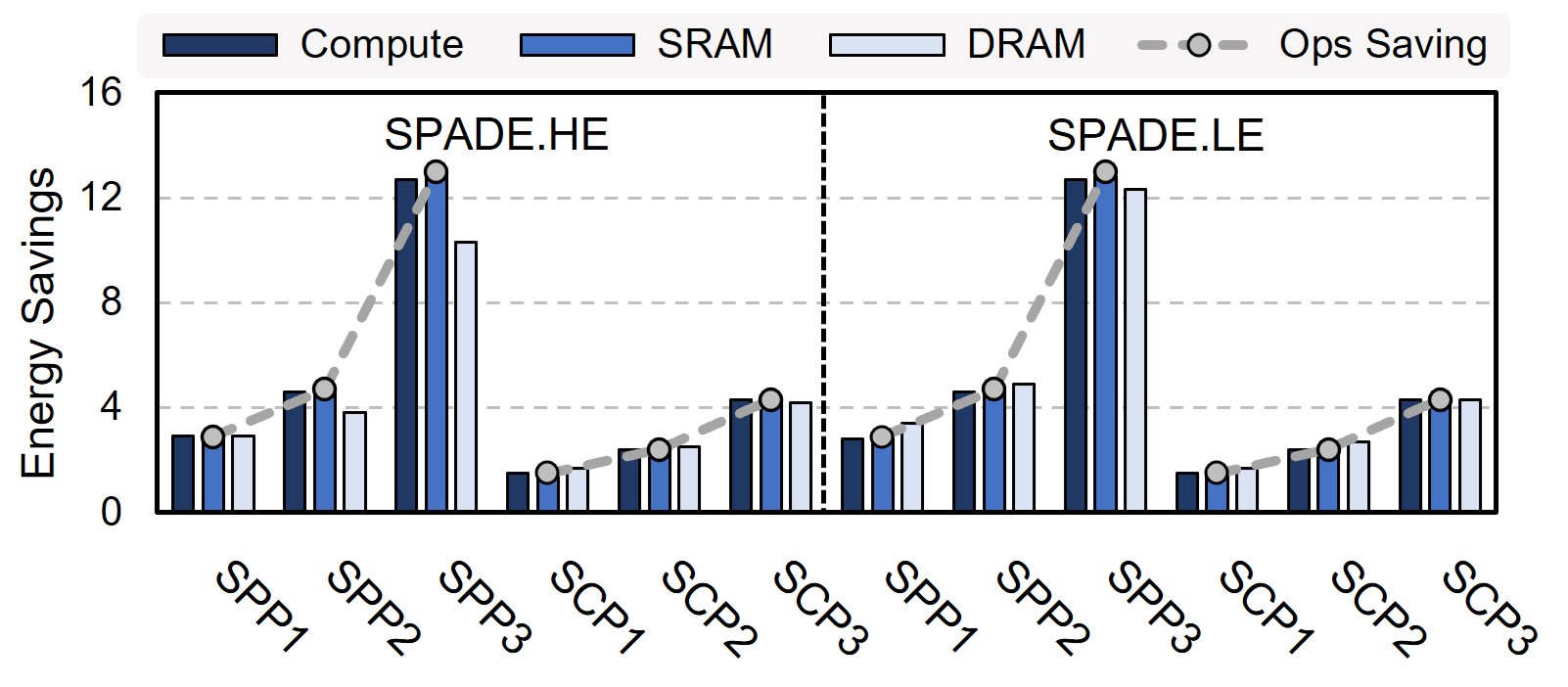}
}
\caption{Breakdown of energy savings of \arch.}
\label{fig:comp_energy}
\end{center}
\vspace{-0.6cm}
\end{figure}

\noindent \textbf{Operation Breakdown of Energy Savings. }
Fig.~\ref{fig:comp_energy} displays the energy savings achieved by each \arch component compared to that of DenseAcc for HE and LE, respectively. Compute and SRAM attain energy savings corresponding to ops savings due to the reduced number of operations. Conversely, DRAM savings slightly lag behind ops savings, particularly for models based on SpConv-S, since the number of outputs generated is for SpConv-S. Nevertheless, the overall trend reveals a strong correlation between ops and energy savings across the various components, demonstrating the effectiveness of \arch's sparsity-scalable performance.

\begin{figure}[t]
\begin{center}
\centerline{
\includegraphics[width=0.87\linewidth]{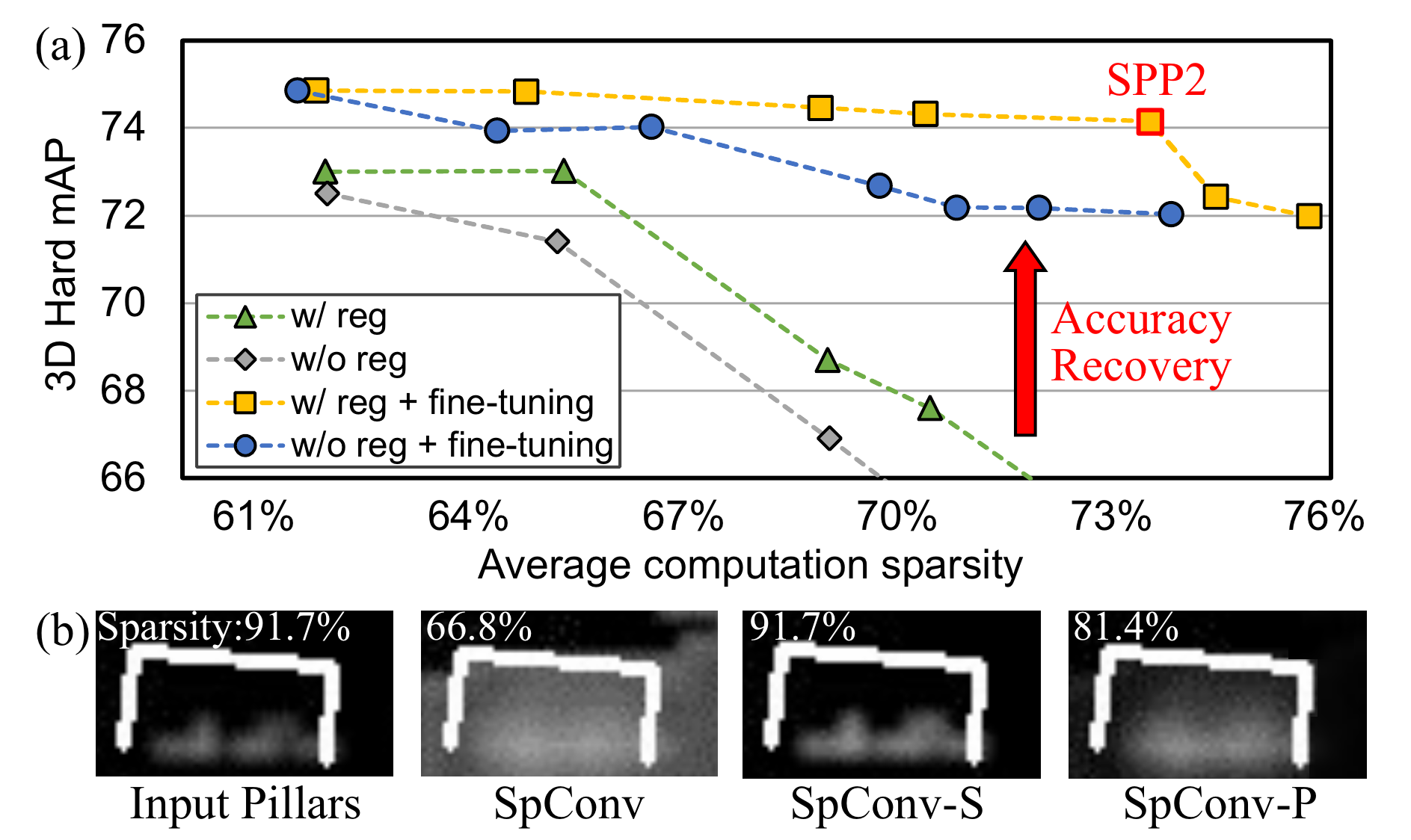}
}
\caption{(a) Accuracy-sparsity trade-off of dynamic pruning with and without regularization and fine-tuning. (b) Feature maps of a single object (\textit{car}) for various sparse convolutions.}
\label{fig:sparsity_accuracy}
\end{center}
\vspace{-0.6cm}
\end{figure}

\noindent \textbf{Accuracy-Sparsity Trade-off of Dynamic Pruning. }
We provide an ablation study on our proposed dynamic pruning method. Fig.~\ref{fig:sparsity_accuracy}(a) depicts the 3D object detection accuracy on the KITTI dataset as sparsity increases, thus reducing the number of operations. Our regularization-based fine-tuning proves particularly beneficial for highly sparse \arch, maintaining baseline accuracy until 26\% sparsity. This allows SPP2 to achieve a four-fold speedup with less than 1\% accuracy degradation. Fig.~\ref{fig:sparsity_accuracy}(b) displays the feature map of a single car object based on different sparse convolution types (white rectangle: ground truth bounding box). The leftmost image shows the initial input pillars, while the images for SpConv, SpConv-S, and SpConv-P display the Stage 1 outputs. Notably, SpConv-S fails to fill the bounding box, SpConv exhibits unnecessary dilations, while SpConv-P strikes a balance, filling the majority of the bounding box without excess.

\subsubsection{Comparison with Point Cloud Accelerator}

To demonstrate the architectural advantages of \arch, we compare it to the leading point cloud accelerator, PointAcc~\cite{lin2021pointacc}. Following \cite{yang2023mars}, we devised a performance simulator for PointAcc where the sorting-based mapping (64-element bitonic merge sorter) is cycle-accurately simulated, and the architecture parameters, such as cache-line size, hit-time, and pipeline latency, are chosen to estimate PointAcc's performance optimistically. We used the same size of memory (PointAcc’s cache, SPADE’s buffer) and MXU (64x64 PE-array). For a clearer comparison, we did not apply dataflow optimization (i.e., there is no overlap of MXU latency and others).

\begin{figure}[t]
\begin{center}
\centerline{
\includegraphics[width=0.97\linewidth]{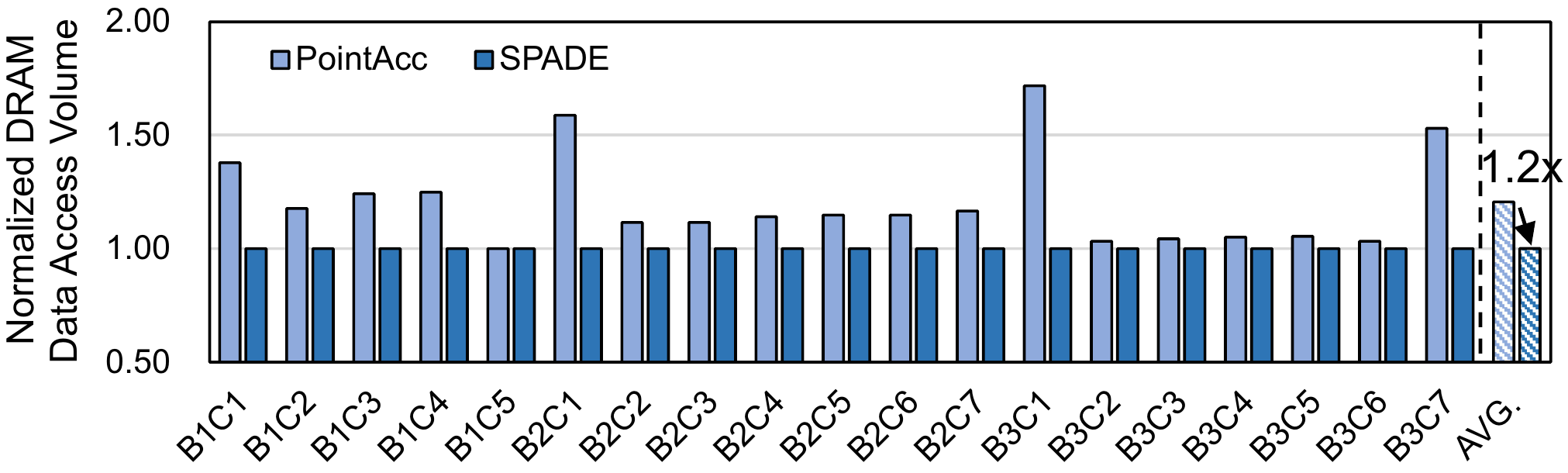}
}
\caption{Comparison of DRAM access volume between \arch and PointAcc~\cite{lin2021pointacc} on SPP2.}
\label{fig:comp_dram_access}
\end{center}
\vspace{-0.9cm}
\end{figure}

\begin{figure}[t]
\begin{center}
\centerline{
\includegraphics[width=0.97\linewidth]{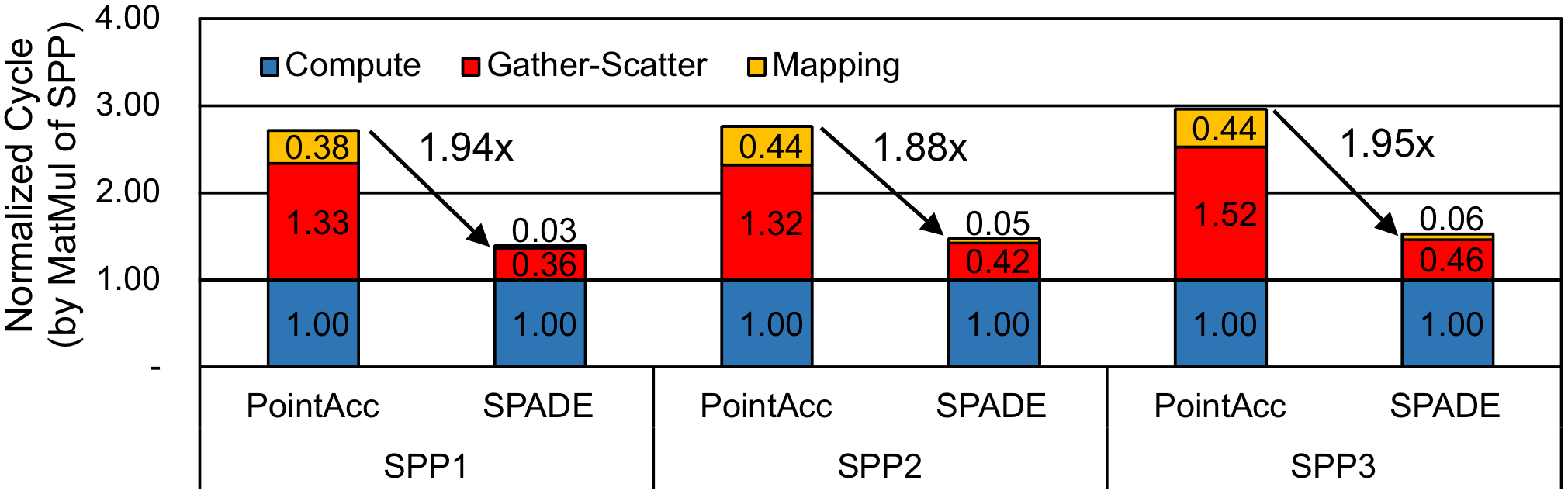}
}
\caption{Latency comparison of \arch with PointAcc~\cite{lin2021pointacc} on various sparse PointPillars.}
\label{fig:comp_pointacc}
\end{center}
\vspace{-0.9cm}
\end{figure}

Our evaluation reveals that PointAcc experiences considerable latency due to mapping and gather-scatter while \arch avoids such inefficiency. Fig.~\ref{fig:comp_dram_access} compares the normalized DRAM accesses measured on SPP1; PointAcc requires, on average, 20\% more DRAM accesses from cache misses. In the latency breakdown of Fig.~\ref{fig:comp_pointacc}, \arch achieves a 1.88-1.95$\times$ speedup thanks to reduced mapping and gather-scatter overheads. The mapping efficiency of SPADE is notably enhanced by RGU, while its gather-scatter efficiency benefits from the sequential data access in GSU, surpassing the improvements from DRAM access reduction. These efficiency gains align with the standalone analysis in Fig.~\ref{fig:rule-gen_archi}(b) and \ref{fig:g_s_archi}(c).

\section{Related Work}
\label{sec:related}

\noindent \textbf{Neural Network Pruning. }
Early neural network pruning methods, such as Optimal Brain Damage~\cite{lecun1989optimal}, identify unimportant weights to be pruned without affecting accuracy~\cite{han2015learning,guo2016dynamic,frankle2018lottery}, but they were not always hardware-friendly. Subsequently, a structured pruning method was proposed~\cite{wen2016learning, louizos2017learning}, forcing hardware-friendly patterns at the cost of accuracy degradation. It has been applied to various domains, including 3D object detection~\cite{stanisz2020optimisation, vemparala2021pruning, zhao2021brief}. In contrast, our activation pruning is dynamic and cannot be done statically. Prior work~\cite{georgiadis2019accelerating, kurtz2020inducing} focused on driving more individual activations to zeros via regularized ReLU, whereas we propose to enforce structural sparsity along the pillar.

\noindent \textbf{Sparse Point Cloud Acceleration. }
Prior work saved computation, exploiting fine-grain sparsity with dedicated accelerator architectures~\cite{han2016eie, parashar2017scnn, rhu2018compressing, sen2018sparce, pal2018outerspace, qin2020sigma, zhang2020sparch, chang2019vscnn}. However, PC-based deep learning models with unique characteristics make processing sparse data challenging~\cite{graham2015sparse, graham2017submanifold, graham20183d, yan2018second, tang2022torchsparse}, and 
recent PC accelerators have attempted to reduce the burden of processing sparse data~\cite{xu2019tigris, feng2020mesorasi, lin2021pointacc, feng2022crescent}. However, none of such accelerators have specifically targeted pillar-based 3D object detection, like PointPillars~\cite{lang2019pointpillars,yin2021center}. Specifically, its intrinsic vector sparsity causes underutilization and bank conflicts in the existing SpConv-Acc and leads to inefficient mapping and cache misses in prior PC accelerators.

\noindent \textbf{Dataflow Optimization. }
Dataflow optimization is crucial for executing neural network operations on dedicated accelerators, and it was explored by numerous proposals~\cite{venkataramani2019performance, yang2020interstellar, parashar2019timeloop, kwon2020maestro, lu2021tenet}. While they primarily focus on dense computation, \cite{wu2022sparseloop} addresses sparsity in weight and activation of neural network operations, targeting individual zero elements. In contrast, ours focuses on the efficient execution of sparse pillars that dynamically change the density of active pillars across layers for each point cloud frame.


\section{Conclusion}
\label{sec:conclusion}
In this paper, we presented \arch, an innovative algorithm-hardware co-design strategy that leverages vector sparsity in pillar-based 3D object detection to dramatically reduce computational demands and enhance performance in autonomous driving perception pipelines. By recognizing and addressing the overlooked aspect of inherent sparsity in PointPillars encoding, \arch introduces a dynamic vector pruning algorithm, a novel sparse coordinate management hardware, and sparsity-aware dataflow optimization. These components collectively transform 2D systolic arrays into efficient vector-sparse convolution accelerators, significantly reducing the required computation by 36.3–89.2\% for state-of-the-art 3D object detection networks. The resultant performance benefits are substantial, yielding 1.3–10.9× speed improvements and 1.5–12.6× energy savings over ideal dense accelerators, and even more pronounced gains compared to existing server and edge platforms, promising enhanced perception capabilities for the next generation of autonomous vehicles.

\section*{Acknowledgment}

This work was partially supported by Institute of Information \& communications Technology Planning \& Evaluation (IITP) grant funded by the Korea government(MSIT) (IITP-2024-RS-2023-00253914 and No. 2022-0-00957) and the National Research Foundation of Korea(NRF) grant funded by the Korea government(MSIT) (No. RS-2023-00260527). This work was also partially supported by a grant from PRISM, one of the seven centers in JUMP 2.0, a Semiconductor Research Corporation (SRC) program sponsored by DARPA. The EDA Tool was supported by the IC Design Education Center.


\bibliographystyle{IEEEtranS}
\bibliography{refs}

\end{document}